\documentclass[acmsmall,screen]{acmart}
\usepackage{array}
\usepackage{subfigure}
\usepackage{multirow}
\usepackage{amsmath}
\usepackage{float}
\usepackage{hyperref} 
\usepackage{graphicx}
\usepackage{rotating}
\usepackage{tabularray}
\usepackage{marvosym}
\usepackage[ruled,linesnumbered]{algorithm2e}
\usepackage{amsmath}
\usepackage{soul}
\usepackage{enumerate}
\usepackage{xcolor}
\usepackage{graphicx}
\usepackage{tabularray}
\AtBeginDocument{%
	\providecommand\BibTeX{{%
			\normalfont B\kern-0.5em{\scshape i\kern-0.25em b}\kern-0.8em\TeX}}}

\acmJournal{TAAS}
\acmVolume{0}
\acmNumber{0}
\acmArticle{0}
\acmMonth{0}

\begin{document}
	
	
	\title{StatuScale: Status-aware and Elastic Scaling Strategy for Microservice Applications}
	
	
	\author{Linfeng Wen}
	\email{lf.wen@siat.ac.cn}
	\affiliation{%
		\institution{Shenzhen Institute of Advanced Technology, Chinese Academy of Sciences; University of Chinese Academy of Sciences}
		\country{China}
	}
	
	\author{Minxian Xu}
	\email{mx.xu@siat.ac.cn}
	\affiliation{%
		\institution{Shenzhen Institute of Advanced Technology, Chinese Academy of Sciences}
		\country{China}
	}
	
	\author{Sukhpal Singh Gill}
	\email{s.s.gill@qmul.ac.uk}
	\affiliation{%
		\institution{Queen Mary University of London}
		\country{UK}
	}
	
	\author{Muhammad Hafizhuddin Hilman}
	\email{muhammad.hilman@ui.ac.id}
	\affiliation{%
		\institution{Universitas Indonesia}
		\country{Indonesia}
	}
	
	\author{Satish Narayana Srirama}
	\email{satish.srirama@uohyd.ac.in}
	\affiliation{%
		\institution{University of Hyderabad}
		\country{India}
	}
	
	\author{Kejiang Ye}
	\email{kj.ye@siat.ac.cn}
	\affiliation{%
		\institution{Shenzhen Institute of Advanced Technology, Chinese Academy of Sciences}
		\country{China}
	}
	
	\author{Chengzhong Xu}
	\email{czxu@um.edu.mo}
	\affiliation{%
		\institution{State Key Lab of IOTSC, University of Macau}
		\country{China}
	}
	
	\authorsaddresses{%
		Authors’ addresses: Linfeng Wen, Minxian Xu (Corresponding Author), Kejiang Ye, Shenzhen Institute of Advanced Technology, Chinese Academy of Sciences, 1068 Xueyuan Avenue, Shenzhen University Town, Shenzhen, Guangdong, China, 518055; emails: lf.wen@siat.ac.cn, mx.xu@siat.ac.cn, kj.ye@siat.ac.cn; Sukhpal Singh Gill, Queen Mary University of London, UK; email: s.s.gill@qmul.ac.uk; Muhammad Hafizhuddin Hilman, Universitas Indonesia, Indonesia; email: muhammad.hilman@ui.ac.id; Satish Narayana Srirama, University of Hyderabad, India; email: satish.srirama@uohyd.ac.in; Chengzhong Xu, State Key Lab of IOTSC, University of Macau, Avenida da Universidade, Taipa, Macau, China, 999078; email: czxu@um.edu.mo}
	
	
	\renewcommand{\shortauthors}{Wen et al.}
	
	
	\begin{abstract}
		Microservice architecture has transformed traditional monolithic applications into lightweight components. Scaling these lightweight microservices is more efficient than scaling servers. However, scaling microservices still faces the challenges resulted from the unexpected spikes or bursts of requests, which are difficult to detect and can degrade performance instantaneously. To address this challenge and ensure the performance of microservice-based applications, we propose a status-aware and elastic scaling framework called \textbf{\textit{StatuScale}}, which is based on load status detector that can select appropriate elastic scaling strategies for differentiated resource scheduling in vertical scaling. Additionally, StatuScale employs a horizontal scaling controller that utilizes comprehensive evaluation and resource reduction to manage the number of replicas for each microservice. We also present a novel metric named correlation factor to evaluate the resource usage efficiency. Finally, we use Kubernetes, an open-source container orchestration and management platform,  and realistic traces from Alibaba to validate our approach. The experimental results have demonstrated that the proposed framework can reduce the average response time in the Sock-Shop application by 8.59\% to 12.34\%, and in the Hotel-Reservation application by 7.30\% to 11.97\%, decrease service level objective violations, and offer better performance in resource usage compared to baselines. 
	\end{abstract}


	
	\received{20 February 2007}
	\received[revised]{12 March 2009}
	\received[accepted]{5 June 2009}
	
	\maketitle
	
	\section{Introduction}
	
	Microservices architecture has emerged as a revolutionary approach in building large and complex software systems \cite{b16,b24}. This architecture has gained immense popularity in recent years due to its ability to deliver flexibility, scalability, and resilience to software applications \cite{b26,TAAS-MICROSERVICE}. In microservices architecture, applications are decomposed into smaller, independently deployable services that communicate with each other through Application Programming Interfaces (APIs) \cite{b23}. Each microservice is responsible for a specific business function and can be developed, deployed, and maintained independently, making it easier to scale and manage the system \cite{b9}.
	
	However,  the prevalent adoption of microservices also presents its own unique set of challenges. One of the significant challenges is dealing with sudden bursts of traffic or load \cite{b30}. Bursty load occurs when there is a sudden surge in traffic or requests to a microservice. This surge could be due to a variety of factors, such as a sudden marketing campaign, a popular event, or even a software glitch \cite{b11}. Regardless of the cause, the microservice must be able to handle the increased traffic without experiencing downtime or degradation in performance \cite{edgeaisim}. This can be a daunting task for microservices, particularly when the burst of traffic is short-lived and unexpected \cite{kumar2023qos}. In such situations, microservices need to allocate sufficient resources quickly and efficiently to meet the increased demand, while also ensuring that the system remains stable and available to users.
	
	To address the above challenges, we propose \textbf{\textit{StatuScale}}, a status-aware and elastic scaling framework for microservices. It aims to handle load bursts by predicting the occurrence of workload spikes in a fine-grained manner and ensuring Quality of Service (QoS) at the target level. 
	
	
	StatuScale utilizes both vertical and horizontal scaling strategies to achieve fine-grained resource management.  In vertical scaling, StatuScale utilizes a resistance line within the load status detector (similar to a trendline in the business market)  to identify whether the current status of a microservice is consistently maintained or not.  It then selects the appropriate elastic scaling strategy accordingly. When the load is stable (e.g., below the resistance line), proactive resource scaling techniques, such as machine learning-based methods, can be employed. However, when the microservice load is unstable (e.g., above the resistance line), a conservative resource scaling approach is adopted to ensure that service level objectives (SLOs) are not violated.  This involves using an Adaptive Proportional-Integral-Derivative (A-PID) controller to maintain resource utilization at the target level. For horizontal scaling, a mechanism based on a comprehensive assessment and resource reduction is designed, triggering horizontal scaling and adjustments when the cumulative value or individual value exceeds the specified threshold. Additionally, a cooling-off period is configured within the horizontal scaling strategy to prevent frequent scaling due to workload fluctuations. 
	
	To demonstrate the effectiveness of StatuScale, we conduct experiments based on Alibaba’s realistic traces, and deploy StatuScale on Kubernetes \cite{b25} platform with two typical microservice-based applications (Sock-Shop\footnote[1]{https://microservices-demo.github.io/} and Hotel-Reservation \cite{b24}). The results demonstrate that our proposed approach outperforms three state-of-the-art baselines in terms of average response time (8.59\% to 12.34\% improvement in Sock-Shop and 7.30\% to 11.97\% in Hotel-Reservation) while maintaining resource usage at a stable status.

	The \textbf{  main contributions} of this work are:
	
	\begin{itemize} 
		
		\item [$ \bullet $]We present StatuScale, a status-aware and elastic scaling framework to handle load bursts and scale resources of microservices to reduce SLO violations. 
		
		\item [$ \bullet $]We propose a resistance line and a support line in vertical scaling to detect whether the load is currently in a relatively stable status, and an extended method based on comprehensive evaluation and resource reduction in horizontal scaling. They work collaborate to effectively handle sudden load spikes and maintain resource utilization at the target level.
		
		\item [$ \bullet $]We evaluate the effectiveness and availability of several baselines via realistic trace and testbed. In terms of traditional and novel metrics (correlation factor), the results have shown that significant performance improvement can be achieved  by StatuScale.
		
	\end{itemize}
	
	The rest of the work is structured as follows: Section~\ref{sec:Related Work} discusses the related work in elastic scaling for cloud and microservice applications. Section~\ref{sec:StatuScale} presents the elastic scaling algorithm of StatuScale. Section~\ref{sec:Performance} demonstrates the performance evaluations of the proposed approach. Finally, Section~\ref{sec:Conclusions} concludes the paper and highlights promising future directions. 
	
	\section{Related Work}\label{sec:Related Work}
	
	Elastic scaling in managing cloud applications and microservices is a popular and well-researched topic. The existing elastic scaling strategies can be mainly divided into three buckets: \romannumeral1) threshold-based, \romannumeral2) control theory-based and \romannumeral3) learning-based.
	
	\subsection{Threshold-based Elastic Scaling Strategies}
	
	Threshold-based elastic scaling strategies predefine suitable target thresholds (e.g. utilization) to trigger scaling actions, such as HPA and VPA built into Kubernetes \cite{b25}, which can be applied to the workloads without apparent trends and difficult for prediction. This category of scaling has been widely adopted in both academia and industry. 
	
	Wong et al. \cite{b1} proposed Hyscale to simultaneously achieve vertical scaling and horizontal scaling to ensure high availability. Xu et al. \cite{spe} proposed an algorithm based on resource utilization threshold for adjusting the number of pods for non-periodic loads, and defined a cooling-off period to ensure that replica removal operations are not executed within this time frame, effectively addressing changes in system load. Liu et al. \cite{b2} proposed a fuzzy logic-based method called Fuzzy Auto-Scaler, which can automatically and adaptively adjust the thresholds for web applications. Rossi et al. \cite{Rossi} proposed an auto-scaling strategy based on dynamic multi-metric thresholds, utilizing reinforcement learning (RL) to autonomously update scaling thresholds to meet the performance requirements of cloud-native applications. Pozdniakova et al. \cite{hppa} enhances Kubernetes' HPA by proposing a method to optimize utilization thresholds. By dynamically adjusting thresholds, it ensures performance-based SLO compliance with minimal resource over-provisioning.
	
	The advantages of using threshold-based elastic scaling strategy are their simplicity and efficiency in making scaling decisions. However, due to the limited capability to response to environment changes, threshold-based approach suffers from low resource utilization and SLO violations. For instance, the static thresholds in \cite{b1,spe} make it difficult to adapt to highly variable workloads. The thresholds in \cite{b2} only set limited number of thresholds manually and cannot achieve fine-grained resource configuration. \cite{Rossi,hppa} requires a lot of trial and error to explore the optimal threshold, leading to performance degradation, and workload bursts cannot be handled efficiently via threshold-based approach with predefined scaling actions. 
	
	\subsection{Control Theory-based Elastic Scaling Strategy} 
	
	Control theory-based elasticity scaling strategy is mainly based on the control system theory \cite{b12,b17}, which aims to monitor and adjust the system's load through feedback control mechanism, thus achieving the system's elasticity scaling. Control theory is a mathematical model used to describe the behavior and control of physical systems. In the control theory based elasticity scaling strategy, its model is used to establish a control system, with the load as input and the system resources as output, to dynamically adjust the system resources using the control system, in order to adapt to the environmental changes.  
	
	Baarzi et al. \cite{b3} proposed SHOWAR, utilizing the three-sigma empirical rule to configure resources and make the decision on horizontal scaling using PID controller. Baresi et al. \cite{Adiscrete-time} proposed an auto-scaling technique based on a grey-box discrete-time feedback controller. Bi et al. \cite{mape} proposed a dynamic microservices framework based on the mean absolute percentage error model. It monitors runtime service data, detects anomalies, and proactively adjusts services using a dynamic window-based elasticity method along with fast scaling and slow shrinking strategies. Hossen et al. \cite{b5} proposed PEMA, a lightweight microservice resource manager that used feedback adjustment to find effective resource allocation strategy.  Rzadca et al. \cite{b4} uses Autopilot to configure resources automatically, it utilizes machine learning and exponential smoothing strategy for fine-tuning, while employing meta-algorithms to adjust parameters. 
	
	Elastic scaling strategies based on control theory can dynamically adjust system resources to adapt to changes in workload, making the system more flexible and resilient. However, there are several issues associated with control theory based elastic scaling strategies. Firstly, such as \cite{b3,Adiscrete-time,mape,b5}, they typically utilize feedback mechanisms to adjust resource allocation but often lack the understanding and learning capabilities of application load characteristics. As a result, they may not effectively adapt to specific load patterns of applications. Moreover, such as \cite{b4}, parameter tuning is also challenging, as control parameters need to be adjusted based on the characteristics of the application, which may require significant time and resources. Lastly, such as \cite{b5}, the feedback cycle is long, as sufficient data must be obtained from the system in order to make informed decisions, which can take long time and it leads to delays that undermine system performance. 
	
	\subsection{Learning-based Elastic Scaling Strategies}
	
	For the cloud applications and microservices with clear periodical trends, a learning-based elastic scaling strategy can be used to characterize historical load data, predict future workloads, and analyze resource requirements for timely resource allocation. Podolskiy et al. \cite{ForecastingModels} extensively compared predictive models for adaptive cloud applications, including ARIMA, exponential smoothing, singular spectrum analysis, support vector regression, and linear regression. Ahamed et al. \cite{deepl} explores proactive resource management in cloud services using deep learning for workloads prediction, various deep learning models are evaluated using real-world workload data. However, there is no one prediction method that is suitable for all time series \cite{AutonomicForecasting}, it is necessary to enhance their adaptability and online learning capabilities.

	Xu et al. \cite{b6} proposed esDNN for cloud load prediction, combining multivariate time series prediction and sliding window to improve prediction accuracy. Podolskiy et al. \cite{MaintainingSLOs} proposed a four-step method, which includes data collection, outlier handling, SLO prediction model establishment, and resource constraint derivation, to effectively address SLO-compatible resource allocation issues. Wang et al. \cite{b7} proposed DeepScaling, which consists of three innovative components: workload prediction using Spatio-temporal Graph Neural Network, CPU utilization estimation using Deep Neural Network, and an adaptive auto-scaling policy based on an improved Deep Q Network. Qiu et al. \cite{b8} proposed a fine-grained resource management framework by leveraging support vector machines to detect SLO violations and make decisions to mitigate the violations with reinforcement learning. Zhang et al. \cite{b22} proposed Sinan, which consists of a Convolutional Neural Network (CNN) and a Boosted Trees (BT) model. The CNN have a global view of the microservice graph and be able to anticipate the impact of dependencies on end-to-end performance, and the BT model is used to predict  the probability of QoS violation. Zhou et al. \cite{AAAI} proposed AHPA, which decomposes load into trend, periodicity, and residual components, and different load forecasting methods are adopted for periodic and non-periodic workloads, such as exponential smoothing and regression forests. The fedformer and Quat-former models serve as replacements when data is abundant.
	
	Learning-based elastic scaling strategies has the capability to understand the characteristics of application workloads and can analyze the demand for resources in advance, thereby enhancing system performance. However, it inevitably faces the problem of model error. The accuracy greatly depends on the selection of the model and the characteristics of the load, and its applicability is limited for high-dimensional and complex loads and microservices. If the system's performance relies entirely on a learning model, such as \cite{b6,MaintainingSLOs}, ensuring performance may be challenging. In addition, such as \cite{b7,b8}, at the beginning of execution, the model may suffer from low accuracy due to the lack of sufficient historical data for modeling, making it difficult to efficiently manage resources at the early stage. In model maintenance, if there are changes in application load characteristics, the cost of retraining the model is extremely high, leading to performance degradation. Lastly, \cite{b7,b8,b22} rely on microservices invocation graphs, leading to poor adaptability of the model during migration.

	\begin{table}
		\caption{Comparison of related work.}
		\label{tab:comprison_table}
		\centering
		\resizebox{\linewidth}{!}{%
			\begin{tblr}{
					column{even} = {c},
					column{1} = {c},
					column{5} = {c},
					column{7} = {c},
					column{9} = {c},
					column{13} = {c},
					cell{1}{1} = {r=2}{},
					cell{1}{2} = {r=2}{},
					cell{1}{3} = {c=3}{c},
					cell{1}{6} = {r=2}{},
					cell{1}{7} = {c=4}{},
					cell{1}{11} = {r=2}{},
					cell{1}{12} = {c=2}{},
					hline{1} = {-}{0.08em},
					hline{2} = {3-5,7-10,12-13}{0.03em},
					hline{3} = {-}{0.05em},
					hline{20} = {1-13}{0.08em},
				}
				\textbf{Approach} &   & \textbf{Technique} &  &  &  & \textbf{Performance Metrics} &  &  &  &  & \textbf{Scaling Mode} & \\
				&  & Threshold & Control Theory & Learning &  & Resource Usage & Response Time & SLO Violation\,\&\,Error & Supply\,\&\,Demand &  & Vertical & Horizontal\\

				~~~~~~~ Wong et al. \cite{b1}  &  & $\checkmark$ &  &  &  & $\checkmark$ & $\checkmark$ & $\checkmark$ &  &  & $\checkmark$ & $\checkmark$\\
				~~~~~~~ Xu et al. \cite{spe}  &  & $\checkmark$ &  & $\checkmark$ &  & $\checkmark$ & $\checkmark$ &  &  &  &  & $\checkmark$\\
				~~~~~~~ Liu et al. \cite{b2}  &  & $\checkmark$ &  &  &  & $\checkmark$ & $\checkmark$ & $\checkmark$ &  &  &  & $\checkmark$\\
				~~~~~~~ Rossi et al. \cite{Rossi}  &  & $\checkmark$ &  & $\checkmark$ &  & $\checkmark$ & $\checkmark$ & $\checkmark$ &  &  &  & $\checkmark$\\
		    	~~~~~~~ Pozdniakova et al. \cite{hppa}  &  & $\checkmark$ &  &  &  & $\checkmark$ &  & $\checkmark$ &  &  &  & $\checkmark$\\

				~~~~~~~ Baarzi et al. \cite{b3}  &  &  & $\checkmark$ &  &  & $\checkmark$ & $\checkmark$ & $\checkmark$ &  &  & $\checkmark$ & $\checkmark$\\
				~~~~~~~ Baresi et al. \cite{Adiscrete-time}  &  &  & $\checkmark$ &  &  &$\checkmark$ & \checkmark &  &  &  & \checkmark & \checkmark\\
				~~~~~~~ Bi et al. \cite{mape}  &  &  &$\checkmark$ &  &  & $\checkmark$ & $\checkmark$ & $\checkmark$ &  &  &  & $\checkmark$\\
				~~~~~~~ Hossen et al. \cite{b5}  &  &  & $\checkmark$ & $\checkmark$ &  & $\checkmark$ & $\checkmark$ & $\checkmark$ &  &  & $\checkmark$ & $\checkmark$\\
				~~~~~~~ Rzadca et al. \cite{b4}  &  &  & $\checkmark$ & $\checkmark$ &  & $\checkmark$ &  & $\checkmark$ &  &  & $\checkmark$ & $\checkmark$\\
				
				~~~~~~~ Xu et al. \cite{b6}  &  &  &  & $\checkmark$ &  & $\checkmark$ &  &  &  &  &  & $\checkmark$\\
				 ~~~~~~~ Podolskiy et al. \cite{MaintainingSLOs}  &  &  &  & $\checkmark$ &  & $\checkmark$ & $\checkmark$ &$\checkmark$ &  &  & $\checkmark$ & \\
				~~~~~~~ Wang et al. \cite{b7}  &  &  &  & $\checkmark$ &  & $\checkmark$ &  &  &  &  & $\checkmark$ & $\checkmark$\\
				~~~~~~~ Qiu et al. \cite{b8}  &  &  &  & $\checkmark$ &  & $\checkmark$ & $\checkmark$ & $\checkmark$ &  &  & $\checkmark$ & \\
			    ~~~~~~~ Zhang et al. \cite{b22}  &  &  &  & $\checkmark$ &  & $\checkmark$ & $\checkmark$ &  &  &  & $\checkmark$ & \\
                ~~~~~~~ Zhou et al. \cite{AAAI}  &  &  & $\checkmark$ & $\checkmark$ &  & $\checkmark$&  &  & $\checkmark$ &  &  & $\checkmark$\\
                ~~~~~~~ StatuScale~(this paper) &  &  & $\checkmark$ & $\checkmark$ &  & $\checkmark$ & $\checkmark$ & $\checkmark$ & $\checkmark$ &  & $\checkmark$ & $\checkmark$\\

			\end{tblr}
		}
	\end{table}
	
	\subsection{Critical Analysis}

    We have proposed a hybrid approach (the hybrid model has gradually emerged as a trend \cite{Anadaptive,Chameleon:A}) named StatuScale, which combines control theory-based and learning-based methods. In Table~\ref{tab:comprison_table}, we compared our proposed method (StatuScale) with the related works discussed above. Differing from threshold-based and control theory-based approaches, StatuScale introduces a lightweight decision tree learning model. This model undergoes rapid retraining and effectively extracts application load characteristics. By analyzing these features, the system gains a better understanding of application behavior patterns and dynamically adjusts resources as needed to meet changing load demands.
        
    Different from learning-based approaches, the occurrence of load bursts and high-dimensional changes renders learning models ineffective with long self-recovery cycles, significantly impacting system performance. To the best of our knowledge, we are the first to apply trend lines from stock price analysis to cloud workload status detector and define a mathematical method for calculating trend lines. This enables effective utilization of load trend analysis for burst detection, facilitating rapid resource allocation to alleviate performance pressure.
        
    Furthermore, we observe that many related works only consider either horizontal or vertical scaling individually, with few addressing both simultaneously. StatuScale introduces a time-window-based fast response and slow contraction horizontal controller to enhance our resource scheduling framework.
	
	\section{StatuScale: A Status-Based Resource Scheduler}\label{sec:StatuScale}
	
	StatuScale integrates both horizontal and vertical scaling, selecting different resource scheduling methods based on the load status. This section will first introduce the system model and objectives of StatuScale (Section~\ref{sec:System Model}), followed by an introduction to the vertical scaling (main part) and horizontal scaling of StatuScale (Section~\ref{Vertical Scaling} and Section~\ref{Horizontal Scaling}), and how they collaborate (Section~\ref{Collaborative Work}).
	
	\begin{figure}[H]
		\centerline{\includegraphics[width=0.95\linewidth]{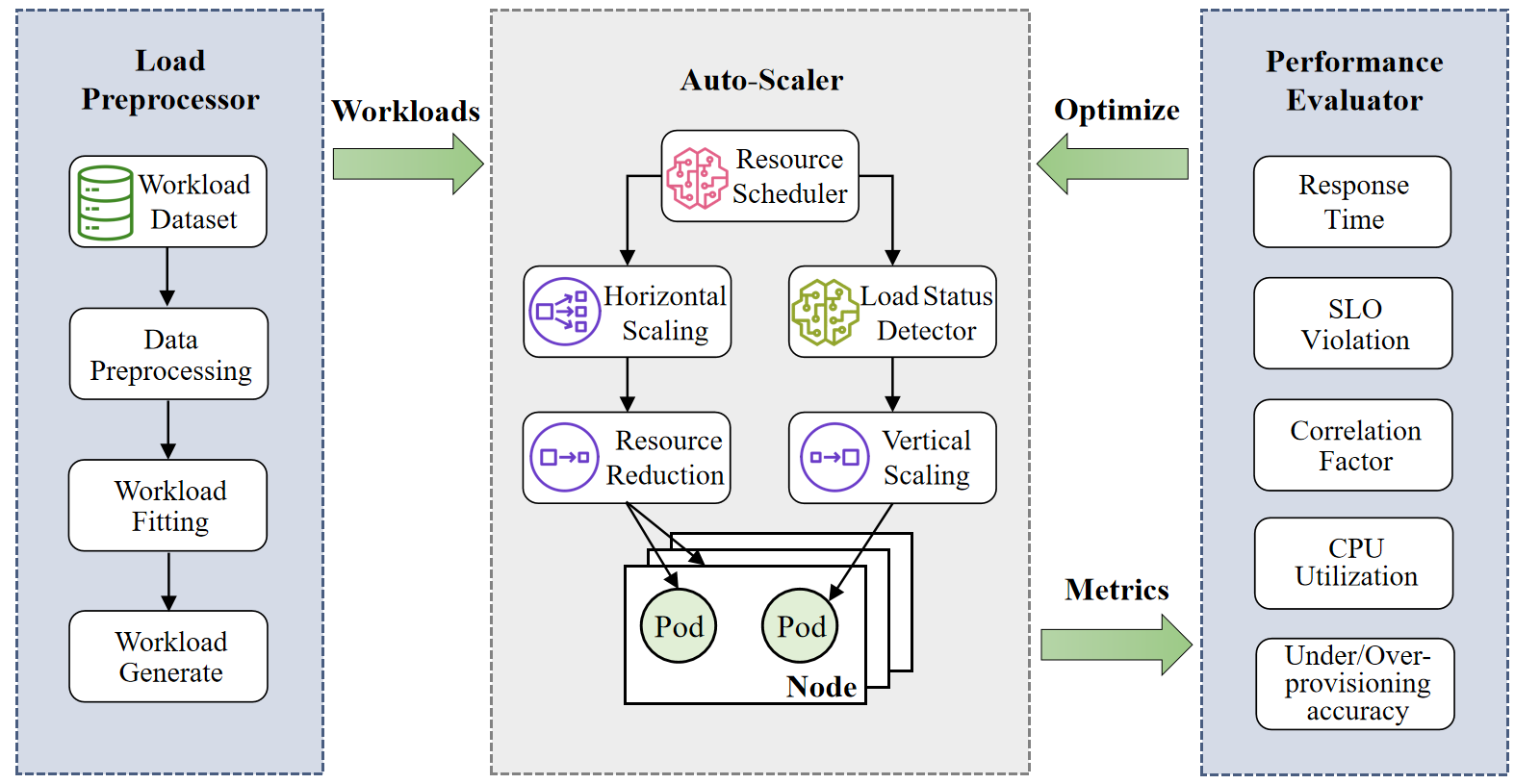}}
		\caption{The system model of StatuScale.} \label{fig:system_model}
	\end{figure}

	\subsection{System Model and Objectives}\label{sec:System Model}
	
	The design goal of StatuScale is to offer an efficient resource management approach for service providers to scale resources. As shown in Fig.~\ref{fig:system_model}, the system model of StatuScale mainly consists of three components: \textit{Load Preprocessor}, \textit{Performance Evaluator} and \textit{Auto-Scaler}. In the following sections, we will describe the design of each component and how they work collaborate. The optimization objective of StatuScale is formulated in Eq.~(\ref{e1}):
	
	\begin{equation}
		\label{e1}
		\begin{split}
			\min  &  \displaystyle  \sum_{{m} \in {M}} \displaystyle{{P}_{{m}}}/{{A}_{{M}}}    \times \sum_{{p} \in {P}}  \displaystyle{{R}_{{p}}}/{{A}_{{P}}} 
			+  \displaystyle \omega^t \sum_{{m} \in {M}}  \displaystyle{{RT}_{{m}}}/{{A}_{{M}}},\\
			& s.t.\; P_m \geq 1 ,\; R_p,RT_m \ge 0 ,\;A_P \geq A_M \ge 0 .
		\end{split}
	\end{equation}
	
	where $M$ represents the set of all microservices in the application, $P_m$ represents the number of pods with microservices $m$ (horizontal quota), $A_ M$ represents the number of microservices in the application, $P$ represents the set of all pods in the application, $R_P$ represents the resource allocation amount (vertical quota) with pod as $p$, $A_P$ represents the number of pods in the application, ${RT}_ m$ represents the response time of microservice $m$. Here parameters $\omega^t$ is a weight that balances resource allocation and performance. The objective of the optimization equation is to minimize the values of both resource quota and response time simultaneously, ensuring the maximization of resource utilization efficiency.
	
	\subsection{Vertical Scaling Controller Based on Load Status Detector}\label{Vertical Scaling}
	
	The change of cloud workloads is influenced by various factors, resulting in periods of stability and turbulence. There may be extended periods of stability when business demands are relatively consistent or when the system is running smoothly. However, due to sudden increases in user demand, such as during specific promotions or product launches, cloud workloads may experience significant changes and become more turbulent. Therefore, for the management and optimization of cloud workloads, dynamic monitoring and analysis are required to understand their changing trends and patterns.
	
	\begin{figure}[H]
		\centerline{\includegraphics[width=\linewidth]{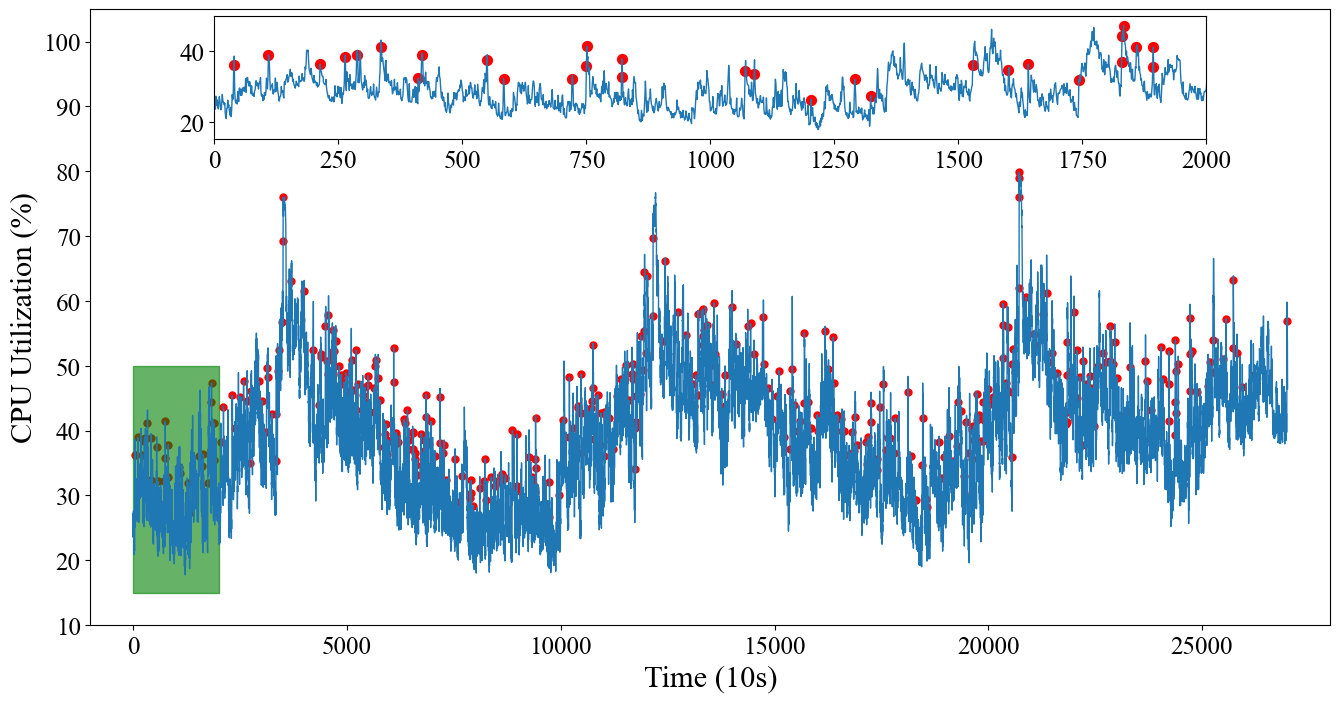}}
		\caption{Underestimating the load leads to a decrease in performance.} \label{fig:unestimate}
	\end{figure}
	
	\subsubsection{Load Predictor based on LightGBM}\label{LightGBM}
	
	The load prediction model provides reliable load forecasting support for cloud service providers. Accurate load prediction and analysis can improve the system's reliability, stability, and performance, thereby increasing the efficiency and availability of the entire system. In order to efficiently make predictions and support rapid scaling decisions, we avoid using heavyweight machine learning or deep learning methods and instead consider adopting the Gradient Boosting Tree algorithm. This algorithm is based on the concept of ensemble learning, where multiple weak learners (typically decision trees) are combined to build a powerful ensemble model. Some well-known gradient boosting trees, such as LightGBM \cite{LIGHTGBM}, have been widely used, with many Kaggle data mining competition winners using it for its excellent performance in most regression and classification problems\footnote[2]{https://towardsdatascience.com/boosting-showdown-scikit-learn-vs-xgboost-vs-lightgbm-vs-catboost-in-sentiment-classification-f7c7f46fd956}, so we use LightGBM as our Load Predictor in StatuScale.
	
	Nonetheless, the challenge of load forecasting stems from its inherent inaccuracy, which arises from the multitude of variables and the unpredictability of future events. Despite leveraging advanced algorithms and data analysis techniques, load forecasting cannot achieve absolute accuracy, we used the Alibaba dataset (refer to Section ~\ref{sec:Configurations} for more details) and LightGBM for load forecasting and found that many times the load predictions were inaccurate, this inherent uncertainty remains a key challenge in load forecasting. As shown in Fig.~\ref{fig:unestimate}, we have already marked the moments of underestimating the load with red dots, most of which are caused by a sudden increase in the load and the system's load is already relatively high. In such cases, it is too late to scale resources, so it is necessary to identify instances of load underestimation in advance in order to take timely actions.
	
	\subsubsection{Load Status Detector}\label{sec:Load Status Detector}
	
	In order to identify potential instances of inaccurate load forecasting and optimize system performance and improve user experience, StatuScale introduces the concepts of resistance and support lines to effectively detect whether the workload is in a "stable" status and take appropriate measures to adapt to different workload statuses.
	
	\begin{figure}[H]
		\centering
		\subfigure[]{\label{fig:a}
			\includegraphics[width=0.49\textwidth]{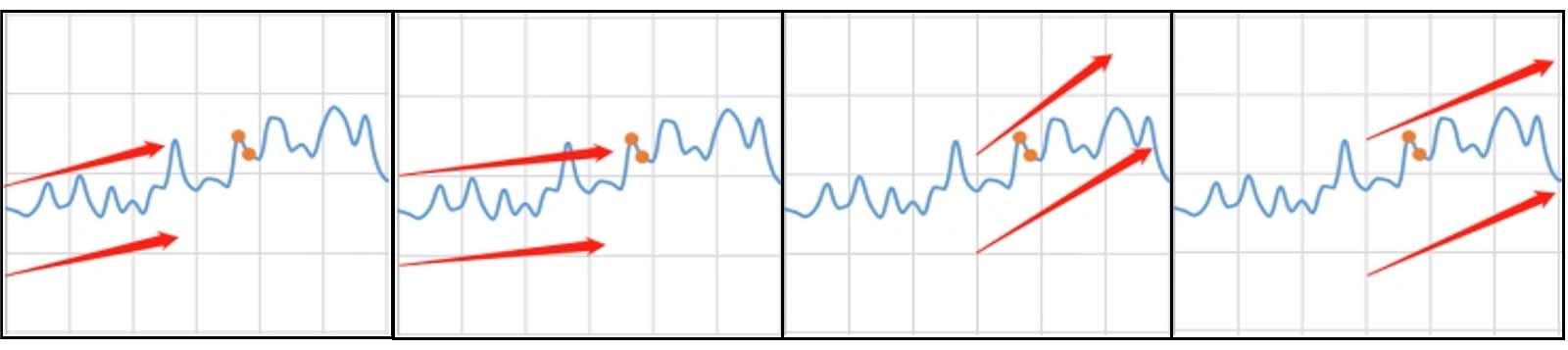}}
		\subfigure[]{
			\includegraphics[width=0.49\textwidth]{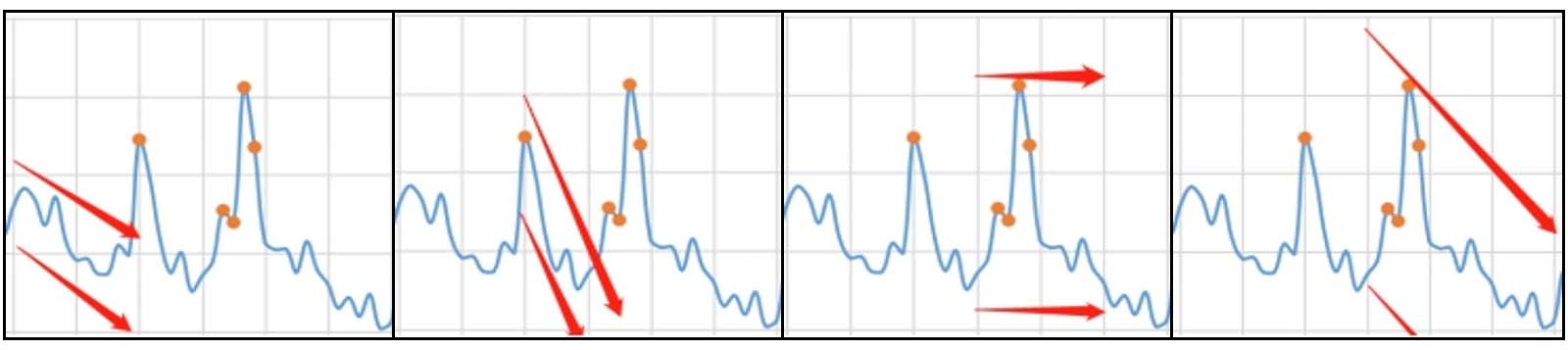}}
		\subfigure[]{
			\includegraphics[width=0.49\textwidth]{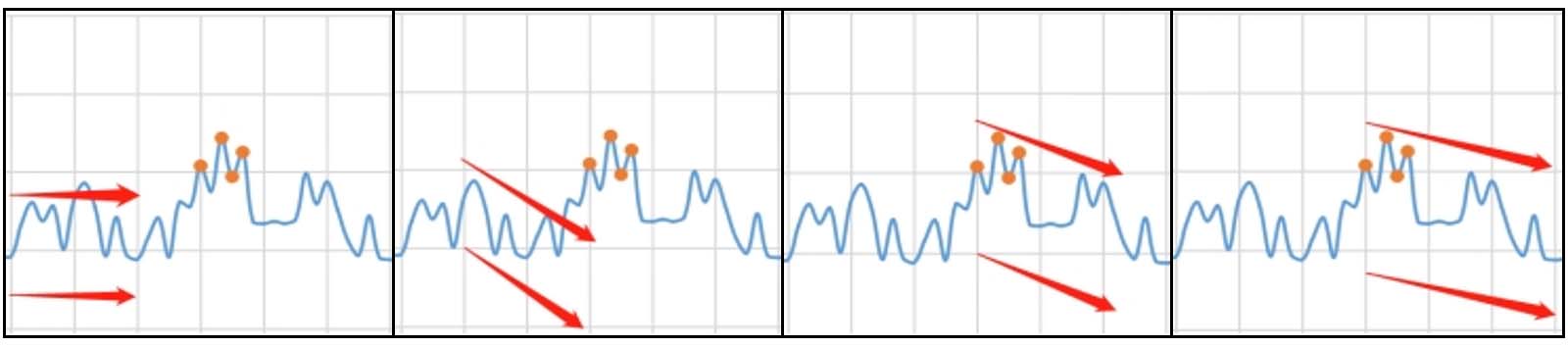}}
		\subfigure[]{
			\includegraphics[width=0.49\textwidth]{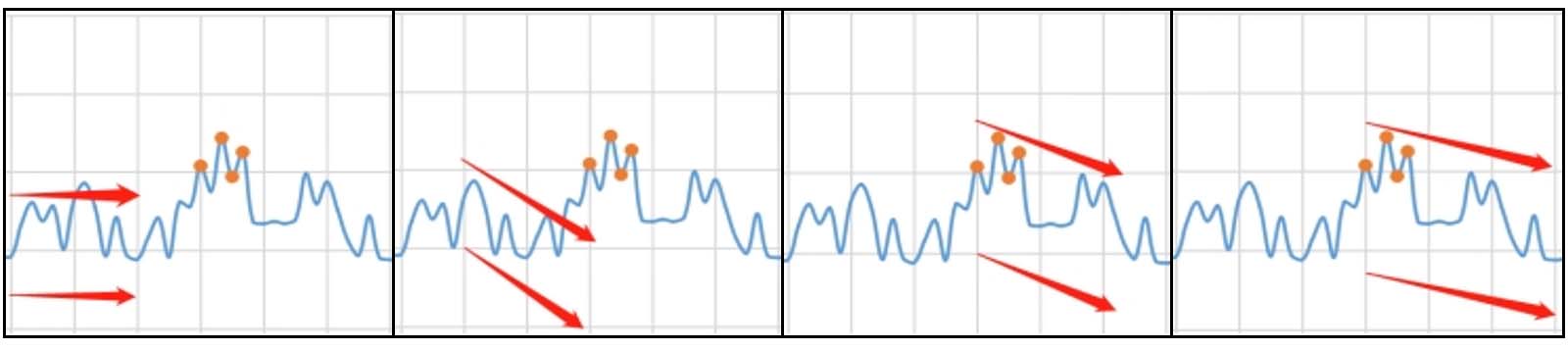}}
		\caption{Examples of resistance line used for load status detector.}
		\label{fig:resistance_line}
	\end{figure}

	Examples of load evaluation using resistance and support lines are shown in Fig.~\ref{fig:resistance_line}, we use Fig.~\ref{fig:a} to illustrate the operation of the resistance line (please note that this figure does not represent actual experimental results). Before assessments, we mark instances of load underestimation, which are indicated by orange dots (as shown in Fig.~\ref{fig:unestimate}). We then divide the x-axis into six segments, and subsequent actions are based on these segments (each segment has 5 data points in StatuScale).
	
	First, we generate resistance and support lines using data from the first segment, and extend them to the next segment. Next, we check whether the data in the second segment exceeds the resistance and support lines or not. If it does, we define it as an unstable load and adopt alternative elastic scaling strategies. If not, we classify it as a stable load, allowing us to continue using the LightGBM-based load prediction elastic scaling strategy. In this example, the data in the second segment does not exceed the resistance and support lines, so we label it as a stable load.
	
Subsequently, we merge the data from the second segment with that of the first segment and update the load resistance and support lines using this combined data. We then extend them to the third segment. As observed in the graph, the load in the third segment has already exceeded the resistance line, so we label it and the next segment (the fourth segment) as an unstable state. The subsequent resistance and support lines data will be generated based on the fourth segment and the process is repeated.
	
	
	Resistance and support lines are applicable to load analysis and can be used to assess load stability. In short, when a load fluctuates between the resistance and support lines, it is considered to be in a relatively stable status. The resistance and support lines represent the boundaries within which the load operates. Within this range, the load may fluctuate up and down, but it generally does not deviate too far from the resistance and support lines. 
	
	Customers prefer platforms that avoid under-provisioning entirely \cite{SPEC}. Therefore, we chose to emphasize how to detect sudden loads by defining a resistance line to prevent adverse effects resulting from under-provisioning, and the definition of a support line is similar and omitted here.
	
	Firstly, we do not consider defining the resistance line as a quadratic or higher-order curve function, because curve functions may result in gradually increasing or decreasing slopes over time, which could lead to the workload "passively" breaking through the resistance line, making overfitting more likely and causing misjudgments. In fact, linear functions are also employed in stock price analysis for similar reasons. However, considering that linear functions are too simplistic and challenging for workload state evaluation, we decided to set them as piecewise linear functions. This design allows for real-time analysis based on the workload's changing conditions, generating workload resistance and support lines. Furthermore, the piecewise linear resistance line can also address the status detector of periodic workloads (such as varying load characteristics during daytime and nighttime), because we set the time window to be dynamic. When the workload is in a stable state (between the resistance and support lines), the time window gradually increases. It adjusts with the workload's variations and is continuously updated. However, when the workloads surpass the resistance line or support line (entering an unstable state), the time window resets to 0, and the resistance line or support line is regenerated. The defined resistance line satisfies Eq.~(\ref{e10}):
	
	\begin{equation}
		\label{e10}
		f(t)=kt+b+\lambda c_v,
	\end{equation}
	
	where $k$ represents the slope of the resistance line, $t$ is time, $b$ is the constant term of the resistance line, and $c_v$ is the coefficient of variation, which is the ratio of the standard deviation $\sigma$ to its corresponding mean value $\mu$ of the data. It characterizes the degree of dispersion of the sample interval and also represents the margin reserved for the resistance line. The greater the degree of dispersion, the larger the allocated buffer space, and vice versa. $\lambda$ is the adjustment parameter for the margin of the resistance line.
	
	The slope and constant term of the resistance line can be determined using polynomial fitting. 
	
	First, a set of fitting data $\left(t_i\ ,\ {Load}_i\right)$ is given, where $i\in\{0, 1, 2, .... , m-1\}$. Then we can make a fitting function $b+kt$, and convert it into a minimum Mean Square Error problem. If there is a set of fitting coefficients that can minimize the  Mean Square Error $\epsilon$, then this set of coefficients can be considered the best. The equation for calculating the Mean Square Error $\epsilon$ is as shown in Eq.~(\ref{e13}):
	
	\begin{equation}
		\label{e13}
		\epsilon=\sum_{i=0}^{m-1}\left(b+kt_i\ -{Load}_i\right)^2.
	\end{equation}
	
	Next, we can take the following two partial derivatives for  $\epsilon$, and set each partial derivative to 0. The results of equations, shown in Eq.~(\ref{e14}), can be solved to determine the values of $k$ and $b$:
	
	\begin{equation}
		\label{e14}
		\left\{\begin{array}{l}
			\displaystyle\frac{\partial \epsilon}{\partial b}=\sum_{i=0}^{m-1} 2\left(b+k t_i-{Load}_i\right)=0, \\
			\displaystyle\frac{\partial \epsilon}{\partial k}=\sum_{i=0}^{m-1} 2 t_i\left(b+k t_i-{Load}_i\right)=0.
		\end{array}\right.
	\end{equation}
	
\begin{figure}[htbp]
		\centering
		\begin{minipage}{0.45\linewidth} 
			\centering
			\includegraphics[width=\linewidth]{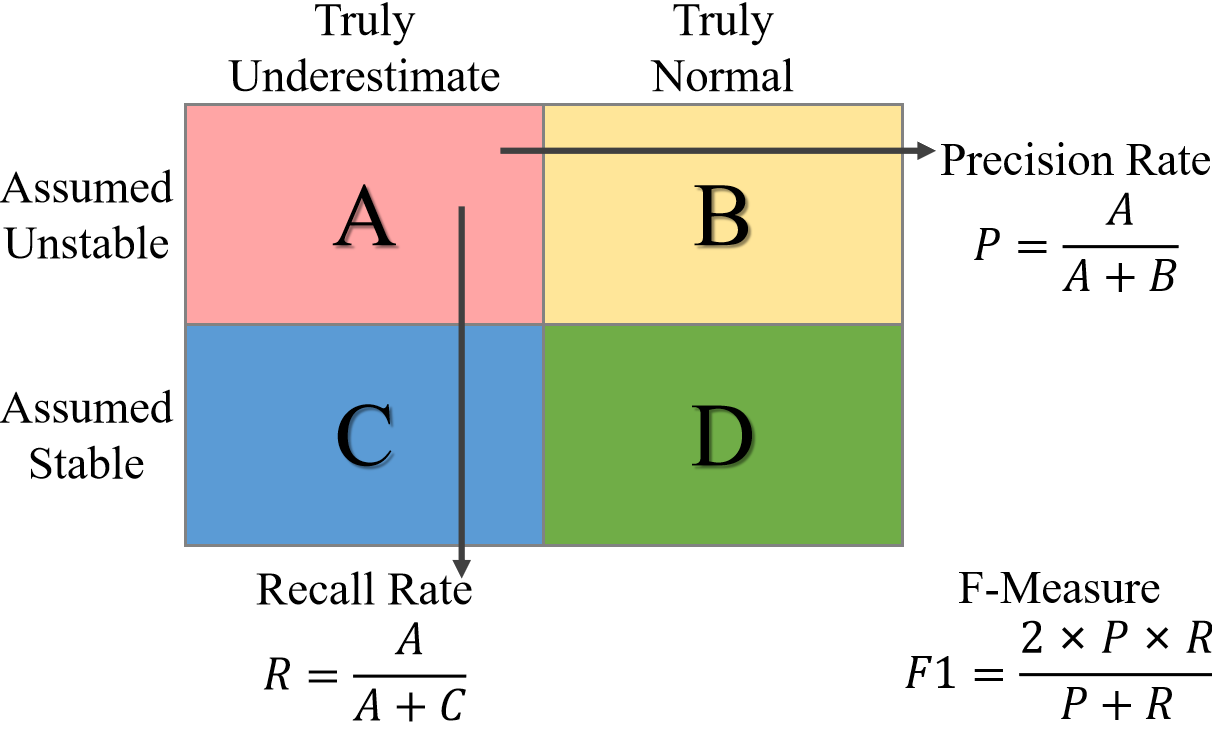}
			\caption{Explanation of Precision and Recall.}
			\label{fig99}
		\end{minipage}%
		\begin{minipage}{0.06\linewidth}
			\centering
			\qquad
		\end{minipage}%
		\begin{minipage}{0.45\linewidth} 
			\centering
			\includegraphics[width=\linewidth ]{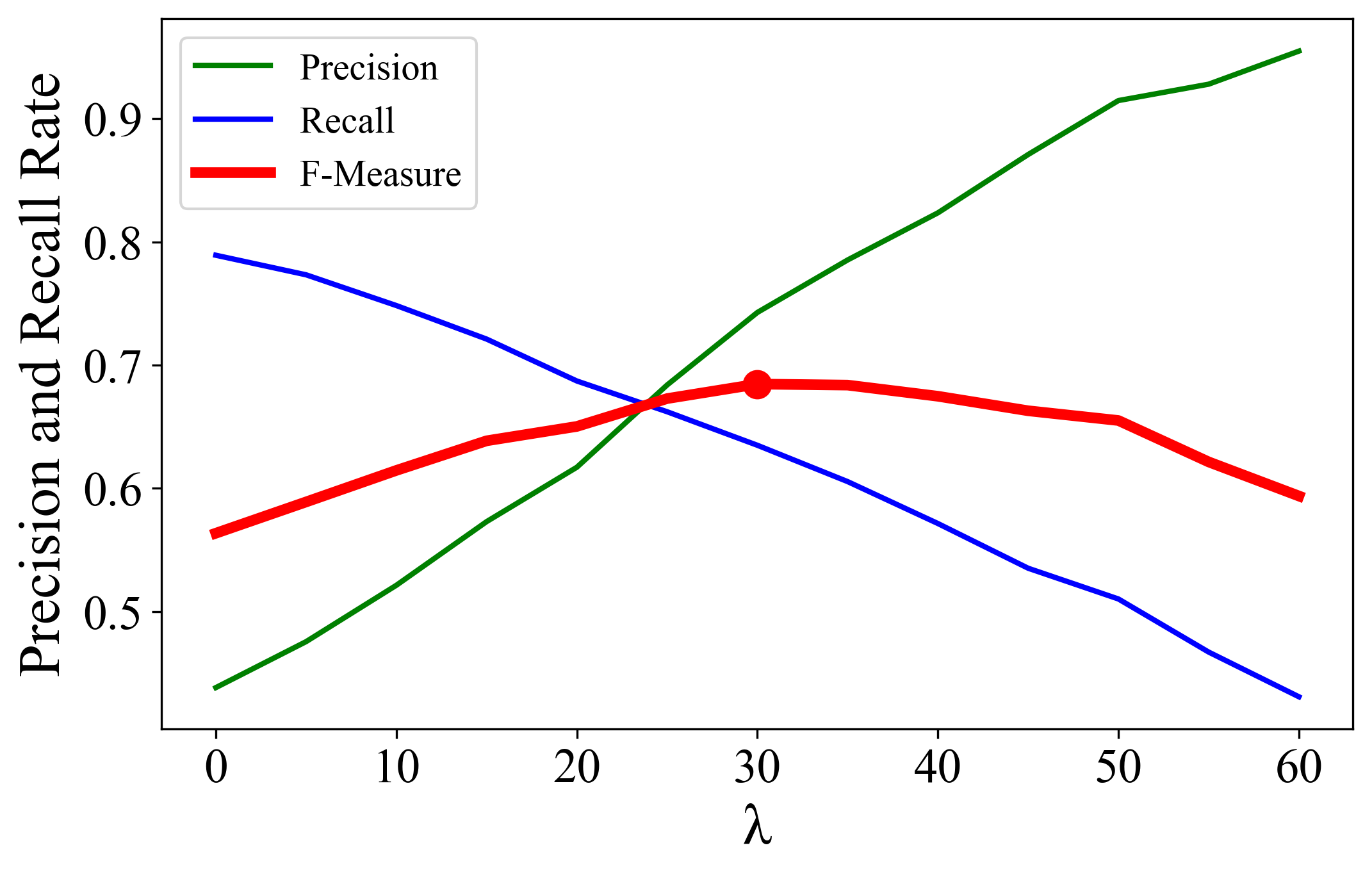}
			\caption{Selection of Parameter $\lambda$.}
			\label{fig7}
		\end{minipage}
	\end{figure}

	In addition, it is necessary to determine the value of the resistance line adjustment parameter $\lambda$. Utilizing the dataset from Alibaba cluster (refer to Section~\ref{sec:Configurations} for details) and conducting simulation experiments with different values of $\lambda$, the suitable parameter $\lambda$ can be chosen based on the experimental results. The problem is defined as follows:
	
	
	
	Precision and Recall are important for evaluating the effectiveness of the Load Status Detector because maintaining a balance between them is critical for assessing and optimizing the detector. As shown in Fig.~\ref{fig99}, we set the number of cases where load is underestimated and detected as unstable as $A$, the number of cases where load is normal and detected as unstable as $B$, the number of cases where load is underestimated and detected as stable as $C$, and the number of cases where load is normal and detected as stable as $D$. We define Precision as $P$, where $P=\frac{A}{A+B}$, it  represents its precision. A low Precision would cause the system to identify many normal states as unstable states, resulting in wastage of resources. We define Recall as $R$, where $R=\frac{A}{A+C}$, it represents the probability of correctly identifying the instances of underestimation. A low Recall would represent that the system fails to identify most moments of underestimation, leading to a performance degradation. Therefore, the objective of a good Detector is to maintain high values of both Precision $P$ and Recall $R$. We use a comprehensive evaluation metric called F-Measure\footnote[3]{https://deepai.org/machine-learning-glossary-and-terms/f-score} for quantification, which is the harmonic mean of Precision and Recall: $F=\frac{2PR}{P+R}$.

	Choosing appropriate parameters is critical for achieving desired results, as parameter selection can have a significant impact on the outcome. Some experiments are conducted using different values of $\lambda$ ($0<\lambda<60$), and the results are shown in Fig.~\ref{fig7}. When $\lambda$ is greater than 60 or less than 0, the Precision or Recall is lower than 0.5. Therefore, we no longer consider cases where $\lambda$ is greater than 60 or less than 0. As the parameter value decreases, the Precision increases while the recall decreases, which is suitable for aggressive resource allocation systems. Conversely, as the parameter value increases, the Precision decreases while the recall increases, which is suitable for conservative resource allocation systems. In conclusion, we recommend selecting the most cost-effective parameter based on the F-measure, and we have marked the maximum value of the F-measure with a dot, corresponding to the case with $\lambda = 30$, where a balance between performance impact and resource usage can be achieved.
	
	\subsubsection{Adaptive Proportional-Integral-Derivative}
	
\begin{figure}[H]
		\centerline{\includegraphics[width=0.85\linewidth]{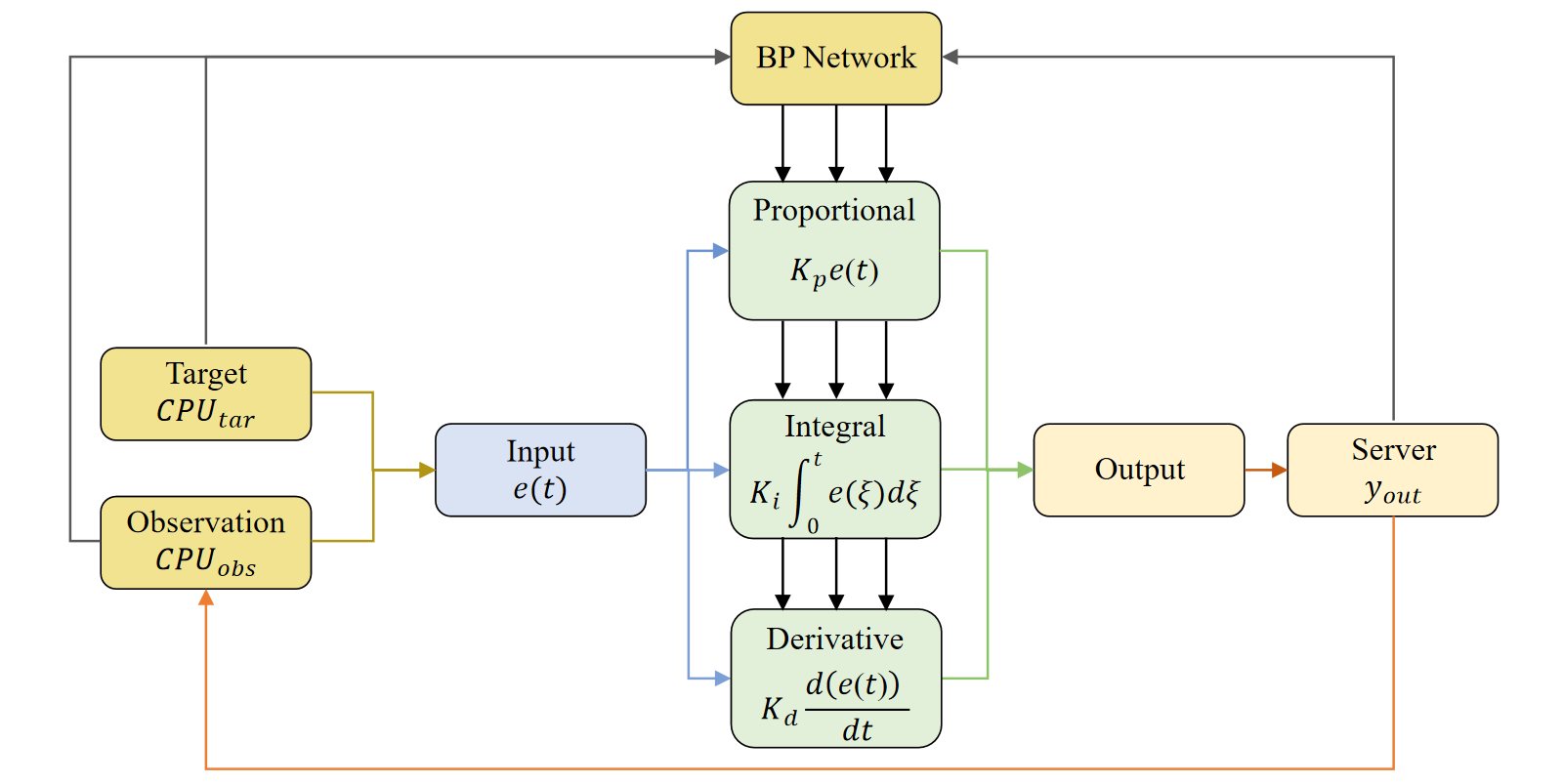}}
		\caption{The architecture of A-PID.} \label{fig5}
	\end{figure}

	For the unstable loads evaluated in Section~\ref{sec:Load Status Detector}, using a Proportional-Integral-Derivative (PID) controller is an effective method for maintaining stability \cite{TAAS-PID}. The PID controller is a widely used feedback controller used in automatic control systems, which can be employed to dynamically adjust system resource allocation to adapt to continuously changing load conditions. The PID controller consists of three components: \romannumeral1) proportional, \romannumeral2) integral, and \romannumeral3) derivative, and its output value is the weighted sum of these three components, as shown in Fig.~\ref{fig5}.
	
	The proportional term is responsible for responding to the current error, and the controller adjusts the output signal based on the size of the error. A large error will result in a large output signal, and vice verse. The proportional term responds quickly but can cause overshoot. The integral term accumulates errors over time and helps to reduce steady-state errors. The derivative term reflects the rate of change of the error and helps to suppress overshoot. By adjusting the weights of each term, the PID controller can balance the response speed, stability, and accuracy to achieve the desired output signal.
	
	StatuScale introduces A-PID for vertical scaling to maintain stable CPU utilization and meet SLO constraints. Unlike traditional PID controllers, A-PID has the capability of parameter self-tuning. Initially, a target value is set and used as the input of the A-PID controller along with the observed value. The output result is obtained by calculating the three proportional-integral-derivative terms and accumulating them, which will be used for resource adjustment of the server. Regarding parameter adjustment, the backpropagation (BP) network is utilized, the target value, actual output value, error value, and bias term are taken as inputs in the input layer. In the middle, there is one hidden layer with 5 neural nodes, and the activation function for the hidden layer is $tanh$. The PID parameters $K_p$, $K_i$, and $K_d$ are outputs in the output layer, and the activation function for the output layer is $sigmoid$, to achieve adaptive parameter adjustment.
	
	\subsection{Horizontal Scaling Controller Based on Comprehensive Evaluation and Resource Reduction}\label{Horizontal Scaling}
	
	Designing a horizontal scaling system is more challenging than a vertical scaling system. Horizontal scaling requires time to create or remove pods, and load balancing is necessary during this period, which may affect service performance and user experience. Additionally, the uncertainty of the workload can result in the system executing unnecessary auto-scaling actions, leading to resource waste.
	
	In the pods co-location context, horizontal scaling adjustments are usually coarse, while vertical scaling adjustments are more fine-grained. Moreover, under light load conditions, vertical scaling performs faster because it can quickly increase system capacity, leading to faster response time and higher steady-state throughput. However, under heavier load conditions, horizontal scaling is more efficient in increasing the system's steady-state capacity, making it more effective overall \cite{ATOM:Model-Driven}.
	
	Therefore, when the workload increases, StatuScale can first assess whether vertical scaling can meet the load requirements or not (vertical scaling are more advantageous in light loads). If not, StatuScale starts with horizontal scaling for a coarse resource adjustment, and then uses vertical scaling for a more precise adjustment (typically employing resource exponential decay \cite{b4}) to meet the application's needs while avoiding resource wastage and cost escalation.

	To overcome these challenges, StatuScale employs a horizontal scaling control mechanism based on comprehensive evaluation and resource reductions. The method involves analyzing and transforming CPU utilization and comparing the transformed results with thresholds to determine whether to perform elastic scaling operations or not. It requires setting upper and lower  thresholds, and determining how CPU utilization will be transformed. The CPU utilization transformed at time $t$ is calculated by Eq.~(\ref{equ:score}), $C_t$ is CPU utilization rate at time $t$, the constant $K$ is a value greater than 1 used to adjust the horizontal scaling of strictness in maintaining target CPU utilization. 
	
	\begin{equation}
		\label{equ:score}
		S_t=\left\{
		\begin{aligned}
			1-{\rm K} ^{CPU_{tar}-C_t} \ ,\  C_t<CPU_{tar},\\
			{\rm K} ^{C_t-CPU_{tar}}-1 \ ,\  C_t \ge CPU_{tar}.\\
		\end{aligned}
		\right
		.
	\end{equation}
	
	When current value $C_t$ is close to the target value $CPU_{tar}$, $|S_t|$ approaches zero. When $C_t$ is far from the target value $CPU_{tar}$, $S_t$ deviates from zero, and as the distance increases, $|S_t|$ increases exponentially. 
	
	To reduce the negative impact of instantaneous bursts, by using a sliding window, the sum of $S_t$ at different time points is computed and compared against upper and lower threshold values to evaluate the need for horizontal scaling. According to Eq.~(\ref{equ:score}), when resource utilization remains consistently high or reaches extremely high levels, horizontal scaling can be triggered immediately to maintain resource utilization within a certain range. In the case of horizontal scaling, the number of replicas to be increased or decreased ($R_n$) is a configurable percentage $\delta$ (defaulting to 10\% \cite{b3}) of the current number of replicas ($R_c$) for the microservice, and the minimum value of $R_n$ is 1. It can be represented by Eq.~(\ref{equ:replica}):
	
	\begin{equation}
		\label{equ:replica}
		R_n=\max(\delta \cdot R_c,1).
	\end{equation}
	
	Despite our careful design, the horizontal scaling system still faces the following issues: when the system detects an increase in load and triggers horizontal scaling, but the scaling has not taken effect yet, the system may misjudge resource inadequacy and trigger horizontal scaling again, leading to multiple scaling operations. This phenomenon also occurs during scaling down. Additionally, during significant load fluctuations, the system may experience frequent scaling operations, potentially destabilizing the system. This frequent operation may increase resource overhead, decrease response speed, or cause service interruptions.
	
    To address this issue, we introduced a cooling-off period to regulate resource scaling. Typically, the length of the cooling-off period can range from a few minutes to tens of minutes, depending on various factors. A longer cooling-off period can reduce the frequency of resource adjustments, thereby lowering the cost of resource adjustment. On the other hand, a shorter cooling-off period can quickly adapt to frequent and intense load fluctuations. However, in our experiments, we set the cooling-off period to 5 minutes based on the practice of existing article \cite{cool}.
	
	Since horizontal expansion involves a relatively large change, fine-tuning of vertical resource reduction is necessary. This involves gradually reducing the vertical resource quota and reclaiming unused resources. The approach defines a decay rate, which reduces the vertical quota by a certain proportion over a specific period of time. This rate, denoted by $\beta$, satisfies $0<\beta<1$, and the vertical quota is reduced by this ratio every period of time $t$. This is expressed as shown in Eq.~(\ref{e9}), $k$ is a constant greater than 1, $V$ is the initial value of the resource.

	\begin{equation}
		\label{e9}
		V(t)=V\cdot k^{\beta^t-1}.
	\end{equation}

	\subsection{Collaborative Work} \label{Collaborative Work}
	
	We combine horizontal and vertical scaling together to make them work collaboratively. Firstly, the load indicator collector uses Kubernetes Metric Server and Prometheus\footnote[4]{https://prometheus.io/} to collect and aggregate resource metric data in the Kubernetes cluster, such as CPU and memory usage. It allows users to access these metric data through APIs, and return the data to our system backend. The collected data is stored in a local database for monitoring and automated horizontal scaling.

		\begin{figure}[H]
		\centerline{\includegraphics[width=0.9\linewidth]{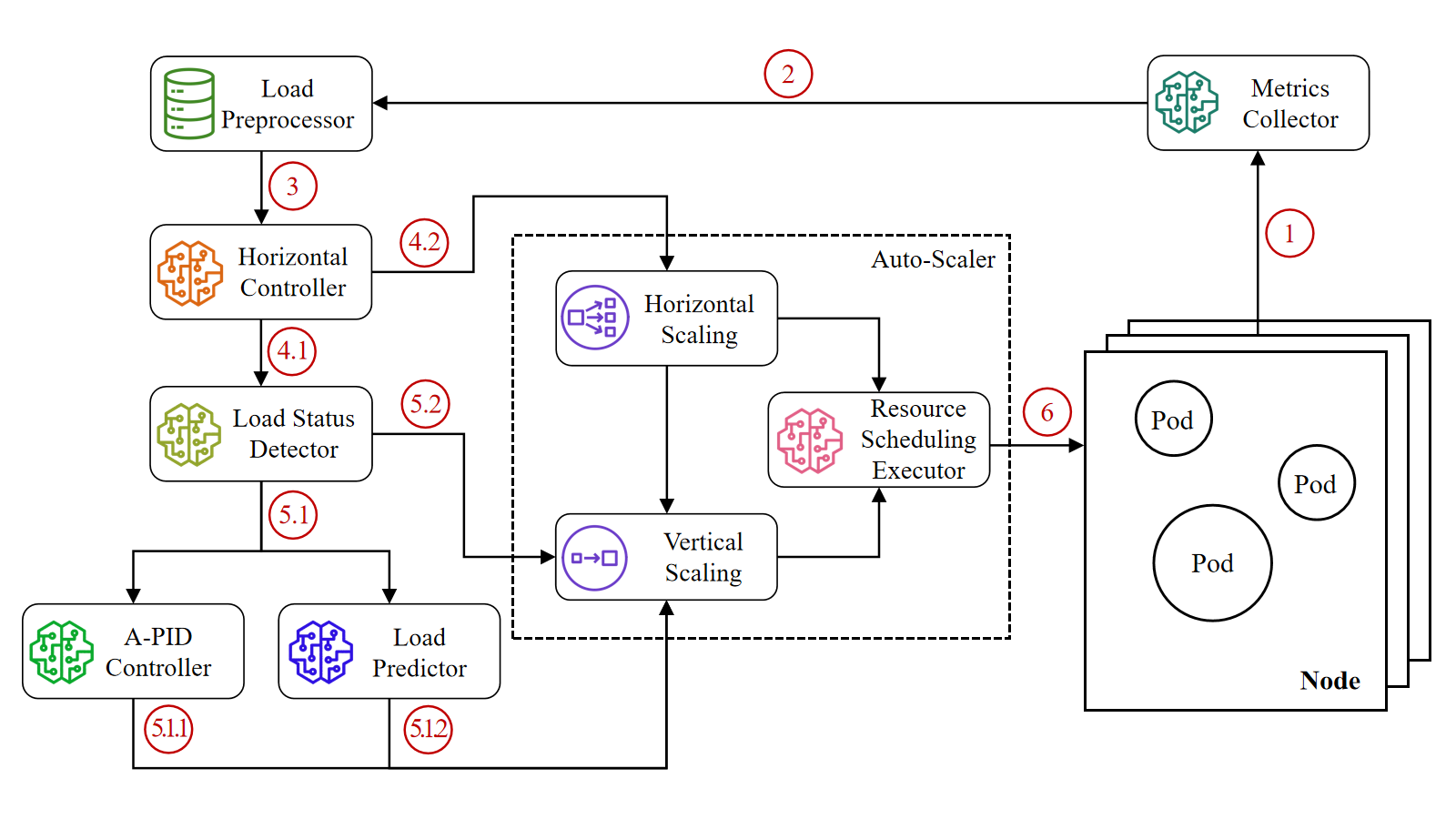}}
		\caption{The execution flow of StatuScale.} \label{fig12}
	\end{figure}


		 \begin{algorithm}
            
		\caption{Elastic scaling algorithm in StatuScale.}\label{algorithm0}
		\KwData{Total upper threshold $S_{ut}$, Single upper threshold $S_{us}$, Total lower threshold $S_{lt}$, Single lower threshold $S_{ls}$, Cooling-off period $T_c$, All collected metrics;}
		\KwResult{Action of scaling;}
		\While{\textbf{True}}
		{	

			Calculate single $S_t$ and total $S_T$ based on Eq.~(\ref{equ:score});

			\eIf{$(S_T>S_{ut}\ or\ S_t>S_{us})\ or\ (S_T<S_{lt}\ or\ S_t<S_{ls})$}{
				Add / Reduce resources based on Eq.~(\ref{equ:replica});
    
				Reduce / Add resources based on Eq.~(\ref{e9});
    
				Suspend for a period of cooling-off period $T_c$;
			}
			{
				\eIf{There is too little load data}{
					Apply vertical scaling strategy based on threshold;
				}
				{
                Generate resistance and support lines based on Eq.~(\ref{e10}), Eq.~(\ref{e13}) and Eq.~(\ref{e14});

                Evaluate load status based on resistance and support lines in Section ~\ref{sec:Load Status Detector};
    
                \eIf{The load is in a stable status}{
						Apply vertical scaling strategy based on predicted workloads via LightGBM in Section ~\ref{LightGBM};
					}
					{Apply vertical scaling strategy based on A-PID in Eq.~(\ref{fig5}) to maintain stability;}
				}
            }
		}
	\end{algorithm}

 	The collected data is sent to the horizontal scaling controller which decides whether horizontal scaling is required based on comprehensive evaluation or not. If horizontal scaling is needed, it is performed first followed by vertical resource adjustments, and a cooling-off period is applied to avoid frequent scaling due to jitter. If horizontal scaling is not performed, vertical inspection is carried out. The data is then analyzed by the load status analyzer. The Load Predictor based on LightGBM is used to forecast the load and allocate resources accordingly. However, if the load is unstable, the A-PID is employed to maintain resource utilization at a stable level. The workflow of StatuScale is shown in Fig.~\ref{fig12}, the numbers on each arrow demonstrates the sequence of executed steps, and the pseudocode of the StatuScale algorithm is shown in Algorithm~\ref{algorithm0}.

 	In StatuScale, horizontal and vertical scaling are conducted as follows:
 	\begin{itemize} 
		
	\item [$ \bullet $]
	
	\textbf{Horizontal scaling}: Kubernetes offers the "kubectl scale" command, allowing adjustment of the replica count for specified resource objects (such as Deployment, ReplicaSet, etc.) via the Linux command line. Thus, horizontal scaling effects can be achieved by scripting automation.
	
 	\item [$ \bullet $]
 	\textbf{Vertical scaling}: Control Groups (cgroups) are a feature of the Linux kernel used to restrict, control, and monitor resource usage of process groups. In Kubernetes, cgroups can be employed to enforce resource limits and management for pods and containers. Hence, vertical scaling effects can be achieved through scripting automation.
 	
 	\end{itemize}
 	
	The time complexity of StatuScale is mainly influenced by the Load Status Detector, the A-PID-based vertical scaling controller, and the LightGBM-based Load Predictor, the time complexity of the horizontal scaling controller is constant, so it can be ignored. Assuming that the number of samples in the time window is $n$, the time complexity of load status detector is $O(n)$. The time complexity of load prediction depends on the LightGBM model, assuming the tree depth is $d$, the number of trees is $N$, and the number of features is $m$. Thus, the time complexity of using LightGBM for prediction is $O(Ndm)$. The time complexity of using A-PID for vertical scaling mainly lies in the forward and backward propagation in the neural network. The time complexity of the forward propagation is related to the size of the neural network and is typically $O(a)$, where $a$ is the number of connections between the input layer and the output layer. The time complexity of the backward propagation is $O(ah^2)$, where $h$ is the number of hidden layers. Therefore, the time complexity of StatuScale is $O(n+\max(Ndm,ah^2))$.

	\section{Performance Evaluations}\label{sec:Performance}
	
	In this section, we  provide a detailed description of the dataset used and the experimental configurations. We  also introduce a new performance evaluation metric, the correlation factor. Additionally, we  conduct experiments on the cluster to compare the performance of StatuScale with several state-of-the-art approaches.  Finally, two sets of experiments were conducted to evaluate the two modules of  StatuScale. The results validate that StatuScale can be effectively applied to automatically optimize cloud resource usage.

	\subsection{ Experimental Configurations}\label{sec:Configurations}
	
	StatuScale is mainly developed using Python 3.9, and resource scaling is performed every 20 seconds. The dataset, microservices, cluster configuration, and baseline methods used in the experiments are as below:

	\begin{itemize} 
		
		\item [$ \bullet $]
		\textbf{Load dataset}: We used the dataset from the Alibaba Cluster\footnote[5]{https://github.com/alibaba/clusterdata}, which provides real production cluster traces. The dataset, named cluster-trace-v2018, was sampled from a production cluster of Alibaba and contains information on approximately 4,000 machines over a period of 8 days, including timestamps, machine IDs, CPU usage, memory usage, network usage, and disk usage. This dataset can accurately represent the workload characteristics of current large-scale cloud clusters. We utilized this dataset as input for workload simulation to evaluate the performance and reliability of applications or systems under various workload conditions.
		
		\item [$ \bullet $]
		\textbf{Microservices demo applications}: The two applications used in our experiments are:
		
		\begin{enumerate}[1)]
			
			\item 
			\textbf{Sock-Shop}: It is an open-source demo application designed to showcase best practices in developing cloud-native applications, which can simulate an online shopping platform and comprises eight microservices, each providing a specific function, such as shopping carts, payments, and inventory.
			
			\item 
			\textbf{Hotel-Reservation}: It is under the microservices architecture and is a distributed system consisting of multiple services. Each service is independent, scalable, replaceable, and communicates via network communication. It includes services such as user, reservation, search, recommendation, and profile.
		\end{enumerate} 
		
		\item [$ \bullet $]
		\textbf{Cluster configuration:} All performance tests were conducted on virtual machines using a Kubernetes cluster consisting of one master and two workers nodes. The operating system used was CentOS-7, with each node having 4 GB of memory and 4 CPU cores.
		
		\item [$ \bullet $]		
		\textbf{Baseline methods:} The three baseline methods used in our experiments are state-of-the-art and representative methods of the three categories of methods in Section~\ref{sec:Related Work} of Related Work.
		
		\begin{enumerate}[1)]
			
			\item
			
			\textbf{GBMScaler} \cite{b27}: It is an elastic scaling strategy based on load prediction, which utilizes LightGBM for model training and elastic scaling based on the predicted results of the model. Choosing the LightGBM-based load prediction as a baseline is justified by the widespread use of this framework in the field of machine learning, its flexibility, high performance, and good tunability and interpretability in load prediction. Importantly, similar to StatuScale, GBMScaler also employs a load predictor based on LightGBM for resource scaling, making it a suitable baseline.
			
			\item 
			
			\textbf{Showar} \cite{b3}: It is a control theory-based elasticity scaling strategy. Since the code for the paper has not been made open-source yet, we have implemented a model based on the ideas from the paper that utilizes a $3\sigma$ empirical formula for vertical scaling without the need for parameter adjustments, and utilizes a PID controller for horizontal scaling, the target values for the PID controller are aligned with StatuScale, and we have changed the metric to CPU utilization. For further optimization, the parameters ($K_p, K_i, K_d$) are dynamically adjusted using a BP neural network. The consistency with StatuScale in using a PID controller for resource scaling further establishes SHOWAR as a suitable baseline.
			
			\item 
			
			\textbf{Hyscale} \cite{b1}: It shares a similar algorithmic concept with Kubernetes' auto-scaler, both utilizing elastic scaling by checking whether the total CPU utilization of all pods on the host exceeds the threshold of the host capacity. But it combines vertical and horizontal scaling to optimize resource utilization and reduce costs. Hyscale has only one adjustable parameter, which is the threshold. We have adjusted the threshold to ensure that the total resource usage quantity across different methods remains consistent in the experimental evaluation. Choosing HyScale as a baseline helps ensure the practicality and applicability of the research, while providing a baseline with generality and reliability for comparison.
			
		\end{enumerate} 
		
	\end{itemize}

	\subsection{Preprocessing and Mapping}
	
\begin{figure}[H]
		\centering
		\subfigure[Profiling with Sock-Shop application.]{\label{fig:profile}
			\includegraphics[width=0.45\textwidth]{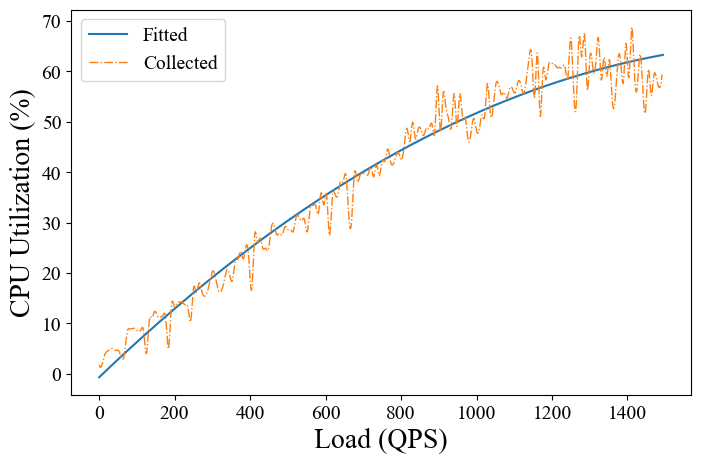}}
		\qquad
		\subfigure[Profiling with Hotel-Reservation application.]         {\label{fig:profile_hotel}
			\includegraphics[width=0.45\textwidth]{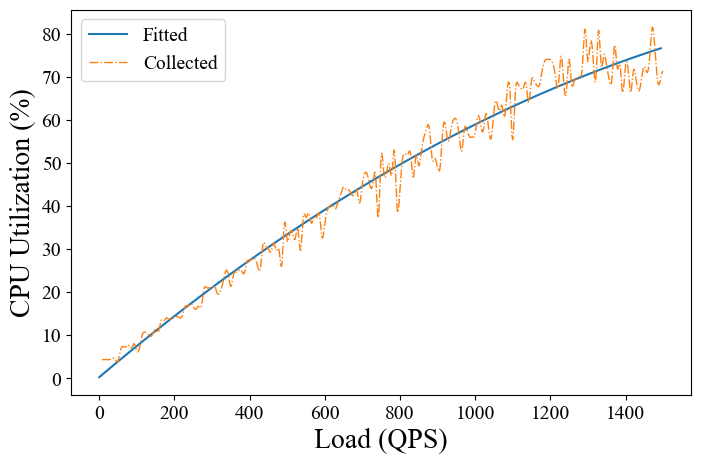}}
		\caption{The relationship between load and CPU utilization.}
		\label{fig0002}
	\end{figure}

	This section covers the preprocessing of raw data, which generates the formatted data that system can process and profile the capability of applications on machines. Even if the loads on two machines are the same, their request processing and CPU utilization may be different due to various factors such as CPU performance, memory and disk performance, task types, other system loads, and operating environments. The differences between machines can be significant, and it is challenging to consider how these factors affect the mapping of loads to CPU utilization. Therefore, we conducted multiple experiments to profile the relationship between loads and CPU utilization, and we use Queries Per Second (QPS) to measure the size of the load. The experimental results based on Sock-Shop are shown in Fig.~\ref{fig:profile}, and the experimental results based on Hotel-Reservation are shown in Fig.~\ref{fig:profile_hotel}. The experimental results are similar to those of previous work \cite{PerformanceModeling}.
	
	After obtaining the mapping relationship between CPU and QPS, we can determine a corresponding QPS based on CPU. We use Locust \footnote[6]{https://locust.io/}, an open-source load testing tool, to simulate a high volume of user requests to evaluate the performance and stability of an application on specific machine. The processed loads can be further analyzed by Load Status Detector to identify the status should be provisioned with scaled resources.
	
	\subsection{Metrics}
	
	In addition to measuring the commonly used metrics such as response time and SLO violations, we referred to some literature on elasticity performance assessment and utilized performance metrics based on the accuracy of supply-demand relationships \cite{SPEC,AnExperimental}. Furthermore, we also innovatively introduced a correlation factor as an additional metric.
	
	\subsubsection{Response time and SLO violations under the same resource budget}
	
	Even under the same resource budget (the resource budget refers to the cumulative product of resource supply within each time unit and each time unit, which is represented as $\int R_t \, dt$, and $R_t$ is the resource provided at time $t$), different elasticity scaling methods can lead to differences in response time and SLO violation rates. We need to evaluate the performance of different elasticity scaling methods from a user perspective. The SLO violation rate is defined as the ratio of the number of violations to the total number of requests.
	
	\subsubsection{Accuracy of supply-demand relationships}

	We are considering evaluating the relationship between resource supply and demand. The resource demand induced by a load is defined as the minimum amount of resources required to achieve a specified performance-related service level objective (SLO) \cite{AnExperimental}. Based on the resource demand and supply curves, we use the following two performance evaluation metrics:
	
	The under-provisioning accuracy metric $a_U$ is defined in Eq.~(\ref{e60}) as the average fraction by which the demand exceeds the supply \cite{SPEC,AnExperimental}:
	\begin{equation}
		\label{e60}
		a_U=\frac{1}{T \cdot R} \sum_{t=1}^T\left(d_t-s_t\right)^{+} \Delta t,
	\end{equation}
	where $T$ is the time period of the experiment expressed in time steps, $R$ is the total number of resources available in the current experimental setup, the resource demand at time $t$ is $d_t$, we calculate $d_t$ based on Fig.~\ref{fig:profile}, we can map any QPS to the corresponding CPU utilization rate. The resource supply is $s_t$, $\left ( x \right ) ^{+} =\max \left ( x,0 \right ) $ is the positive part of $x$, and $\Delta t$ is the time elapsed between two subsequent measurements. Similarly, we define the over-provisioning accuracy $a_O$ as shown in Eq.~(\ref{e61}):
	\begin{equation}
		\label{e61}
		a_O=\frac{1}{T \cdot R} \sum_{t=1}^T\left(s_t-d_t\right)^{+} \Delta t.
	\end{equation}
		
		
		
		
		
		
		
				

	\subsubsection{Correlation factor of supply-demand relationships}
	
	In situations where the total resource supply remains constant, we make the assumption that the supply curve will exhibit a similar trend to that of the demand curve under ideal circumstances. This means that when there is a sudden increase in the load on a microservice, there will be a greater demand for resources, correspondingly, the supply of resources has to increase. If this assumption is not upheld, it will undoubtedly have a negative impact on the performance of the microservice, because the current resource supply will be unable to immediately cope with the unexpected spikes in load, resulting in processing delays and increased response times.
	
	The alignment between the supply curve and the demand curve reflects the efficient utilization of resources. In such scenarios, reducing the allocation of resources will have a relatively minor effect on the microservice, while increasing the allocation of resources can significantly improve resource efficiency. Therefore, the resemblance between the supply curve and demand curve indicates the effectiveness and appropriateness of various resource scheduling methods.

	In CPU-intensive tasks, the supply curve can be represented by the curve showing changes in CPU allocation, while the demand curve can be represented by the curve depicting changes in load. We define a metric called the correlation factor to quantify the similarity between two curves. 	The correlation factor is essentially similar to R-squared, an important statistical metric used to assess the goodness of fit of regression models. It assists us in evaluating the elasticity performance of different methods. However, we did not directly use R-squared here because $d_t$ is collected from Locust, which generates loads periodically, and $s_t$ is collected from Prometheus, which periodically gathers resource usage data. Although their timing tasks have been set to be consistent, due to various reasons, their periods cannot be completely aligned. Therefore, after the tasks are completed, there are issues of inconsistent sample sizes and time drift between the two curves, rendering R-squared inapplicable. 

    Thus, we propose a new method in this paper, mainly inspired by the Dynamic Time Warping (DTW) algorithm \cite{DTW}, commonly used in fields such as speech recognition. It effectively compares time series with time offsets, variable speeds, or different lengths.
	
	There may exist disparities in the number of sample points and time lag between the two curves. However, the DTW algorithm can effectively address this challenge. The fundamental concept is to align the two time series in such a way that their similarities can be compared at various time points. For two time series of length $m$ and $n$, denoted as $X=\left\{x_1, x_2, ..., x_m\right\}$ and $Y=\left\{ y_1, y_2, ..., y_n \right\}$, the DTW algorithm can compute their shortest distance through the following steps.
	
	
	\begin{itemize} 
		
		\item [$ \bullet $]
		To make the two datasets comparable, we transform and scale one set of data such that their mean and standard deviation match the other set of data. Specifically, assuming that the first set of data has an average value of $\mu_X$ and a standard deviation of $\sigma_X$, and the second set of data has an average value of $\mu_Y$ with a standard deviation of $\sigma_Y$, each value $x_i$ in the first set of data should undergo the transformation depicted in Eq.~(\ref{e15}):
		    \begin{equation}
			    \label{e15}
			    {x^\prime}_i=\left(x_i-\mu_X\right)\times\frac{\sigma_Y}{\sigma_X}+\mu_Y,
			    \end{equation}
		where ${x ^ \prime}_i$ denotes the transformed value, and $\frac { \sigma_Y} { \sigma_X}$ represents the ratio of standard deviations between the two datasets. This transformation aims to shift the average of the first dataset to $\mu_Y$ and scale the standard deviation to $\sigma_Y$. Consequently, the first dataset will have the same mean and variance as the second dataset.
		
		\item [$ \bullet $]
		We define a distance matrix $D$ of $m \times n$, where $D_ {i, j}$ represents the distance between $X$ and $Y$ at time points $i$ and $j$. Initialize the distance matrix $D$ so that $D_ {i, j}= \infty$, indicating that two time points cannot be directly connected.
		
		\item [$ \bullet $]
		Starting from the top left corner, the DTW algorithm gradually fills in matrix $D$ and calculates the minimum distance for each position $(i, j)$. Specifically, for position $(i, j)$, it calculates its distance to three positions $(i-1, j-1)$, $(i, j-1)$, and $(i-1, j)$, and selects the minimum value as the current distance. This process can be represented by Eq.~(\ref{e16}):
		\begin{equation}
			\label{e16}
			D_{i,j}=min\left\{\begin{matrix}D_{i-1,j}+d\left(x_i,y_j\right)\\D_{i,j-1}+d\left(x_i,y_j\right)\\D_{i-1,j-1}+d\left(x_i,y_j\right)\\\end{matrix}\right.,
			\end{equation}
		where $d\left(x_i,y_j\right)$ represents the distance between $x_i$ and $y_j$, which can be the Euclidean distance, Manhattan distance, or other distance measures.
		
		\item [$ \bullet $]
		Finally, element $D$ in the lower right corner of matrix $D_ {m, n}$ is the shortest distance between $X$ and $Y$. We can minimize this distance by adjusting the alignment between the two time series. After calculating the DTW distance $D_{m-1,n-1}$, the correlation factor ($CF$) are defined as shown in Eq.~(\ref{e17}), where $\max(m,n)$ represents the maximum sequence length.
		\begin{equation}
			\label{e17}
			CF={\max(m,n)}/{D_{m-1,n-1}}.
			\end{equation}
   
		\end{itemize} 
	
	\subsection{Comprehensive Evaluations}
	
	In CPU-intensive workloads \cite{b13}, insufficient CPU resources allocated to microservices can impact service availability and increase response time, this is an issue that we need to avoid as much as possible. In other words, our experiments aim to reduce response time and SLO violations while maintaining the CPU utilization at target level ($\pm 1\%$). Through the following experiments, we present the performance of StatuScale compared to other baselines, with each experiment repeated five times, and the results presented below are all with average values and 95th percentile confidence interval (CI).
		\begin{figure}
		\centering
		\subfigure[Response time]{\label{fig14}
			\includegraphics[width=0.235\textwidth]{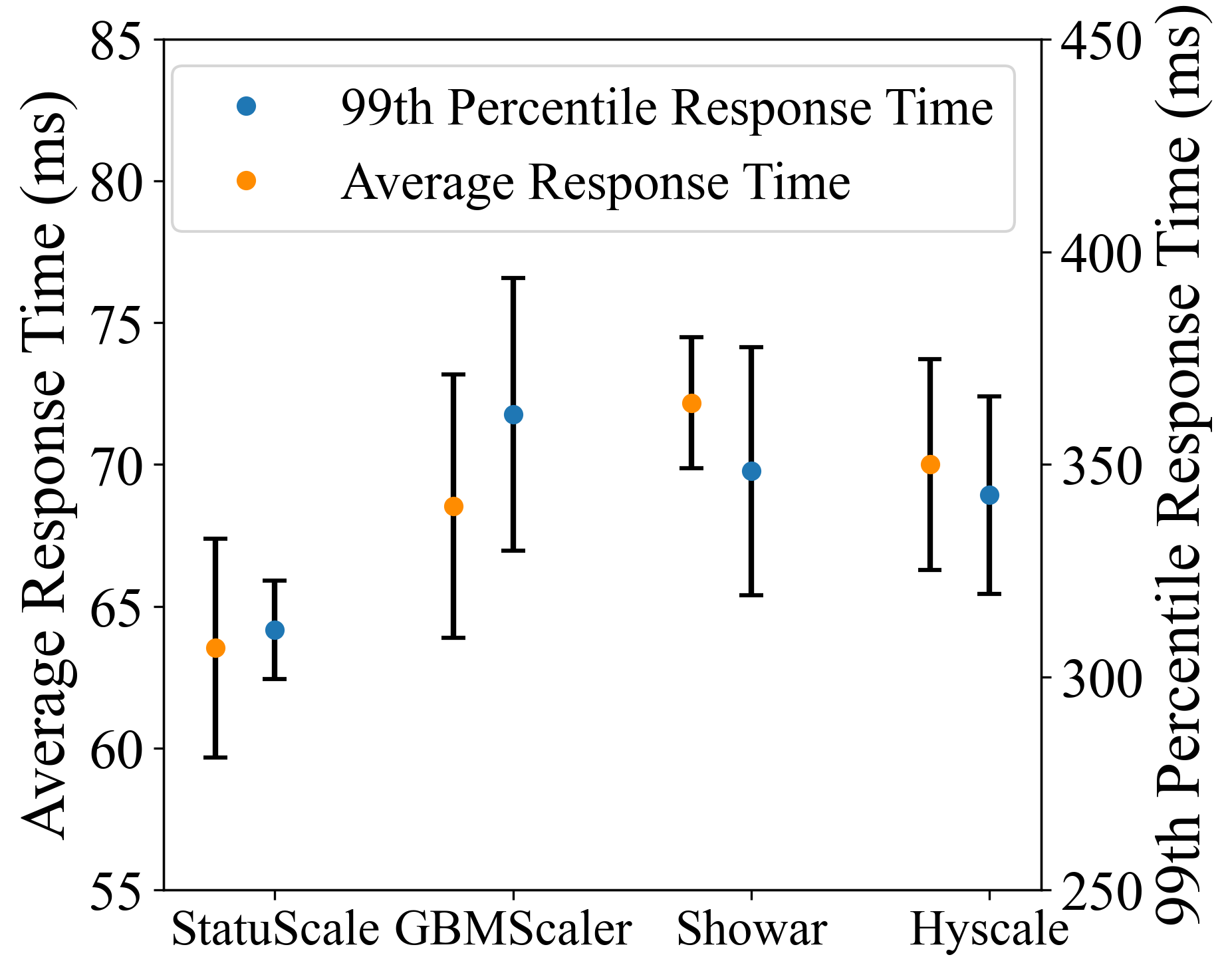}}
		\subfigure[	CDF of response time]{\label{fig15}
			\includegraphics[width=0.235\textwidth]{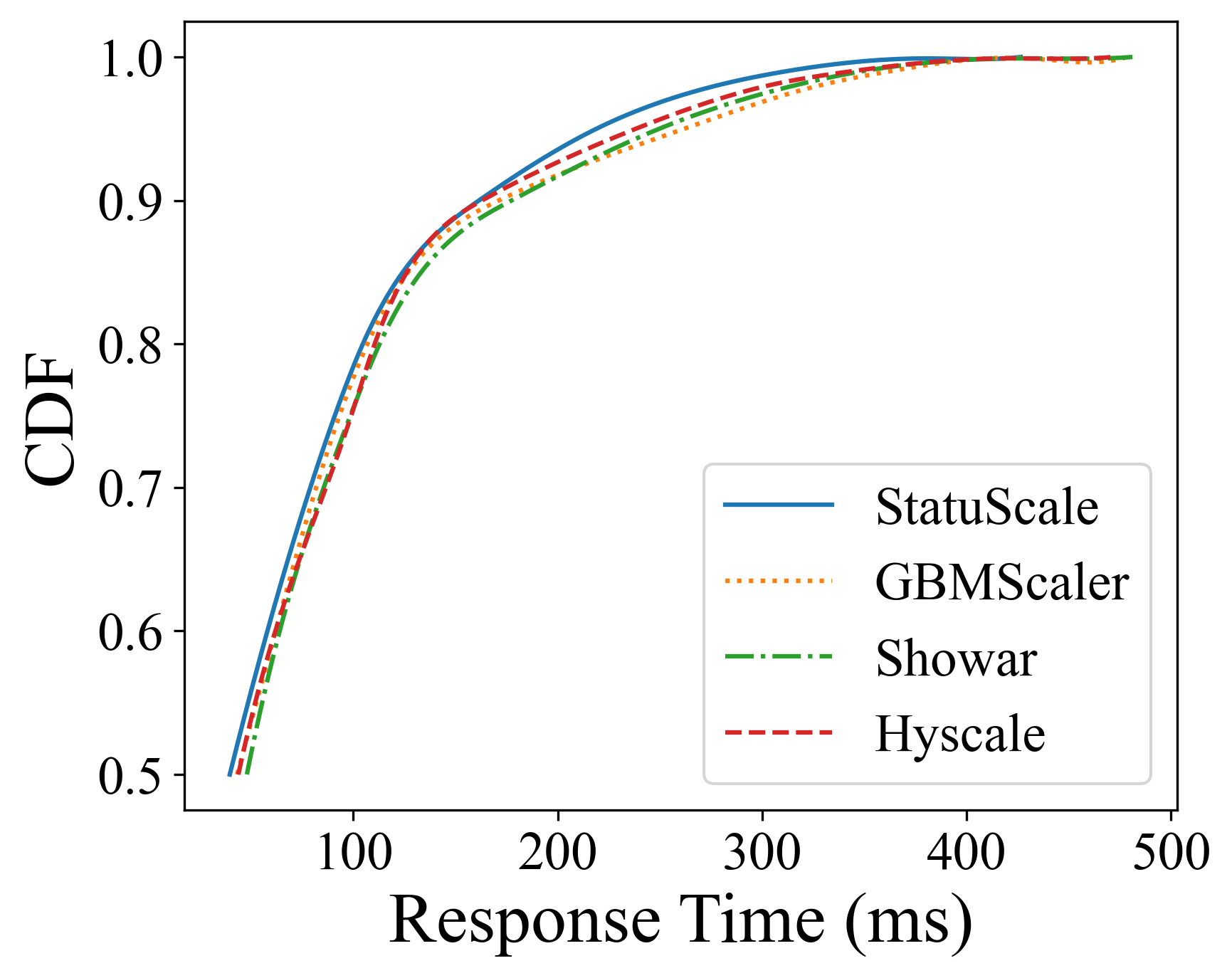}}
		\subfigure[	SLO violations]{\label{fig16}
			\includegraphics[width=0.235\textwidth]{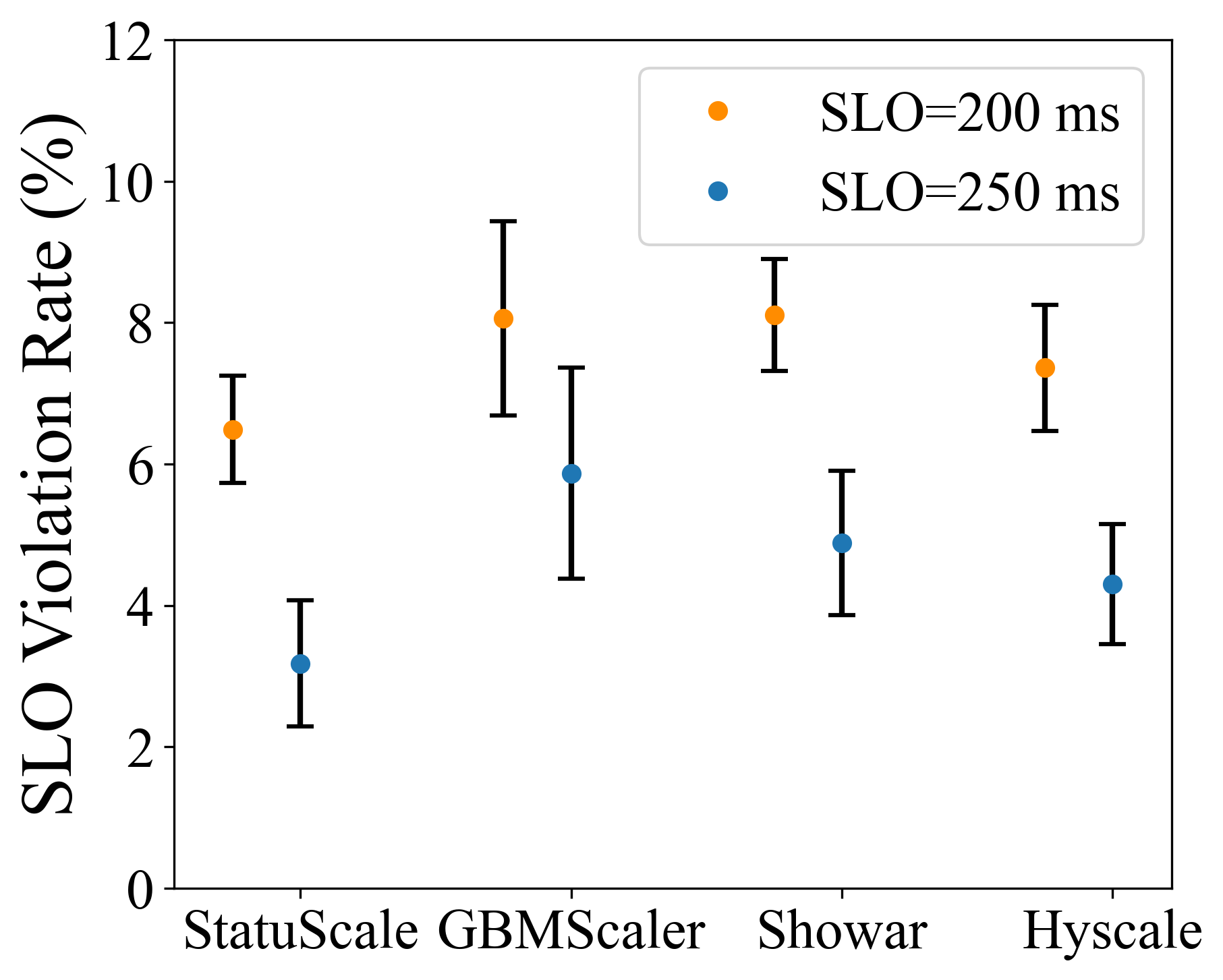}}
		\subfigure[	Correlation factor]{\label{fig21}
			\includegraphics[width=0.235\textwidth]{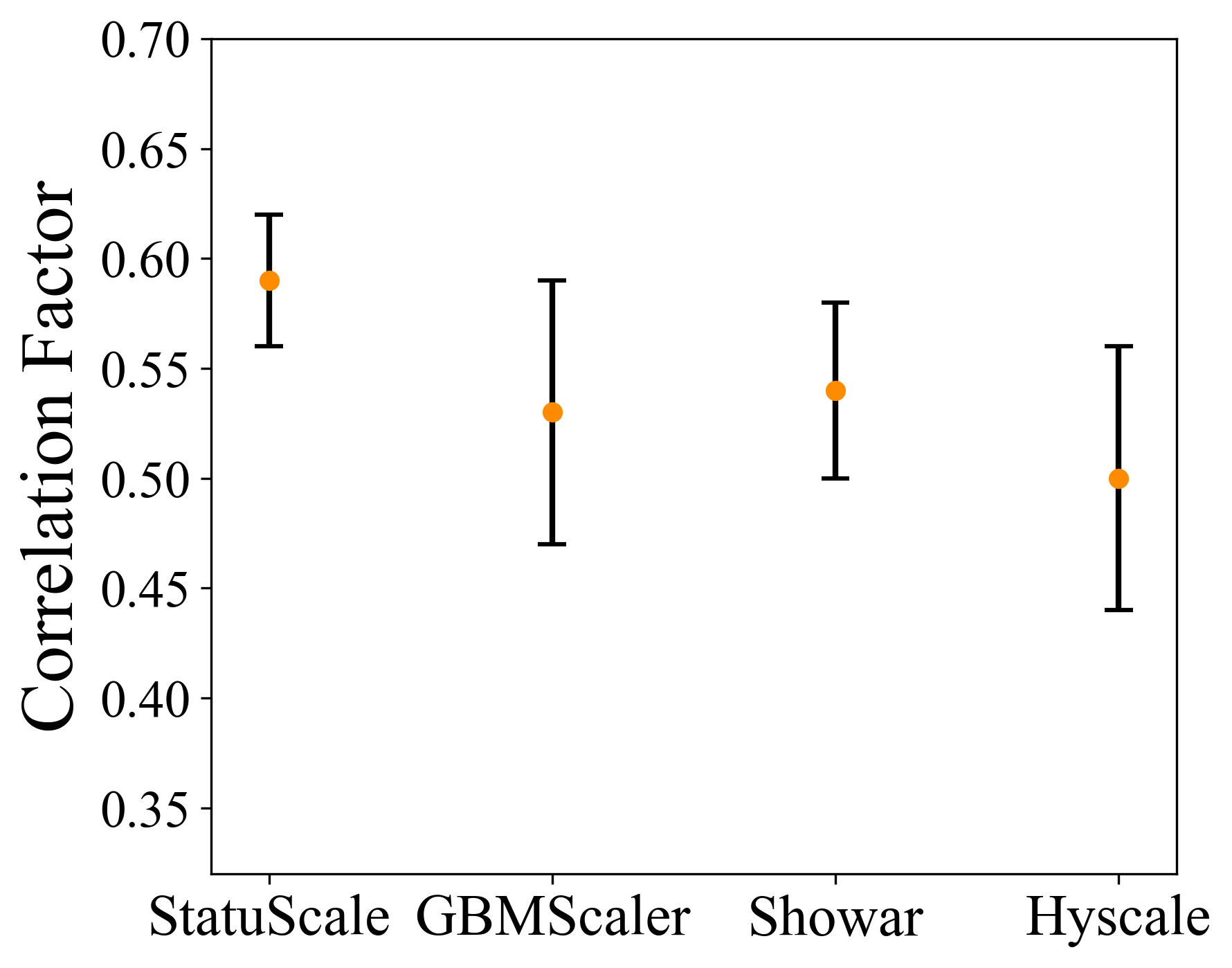}}
		\caption{Performance comparison with Sock-Shop application.}
		\label{fig00}
	\end{figure}
        	\begin{figure}
		\centering
		\subfigure[StatuScale]{\label{fig:subfig:a}
			\includegraphics[width=0.49\textwidth]{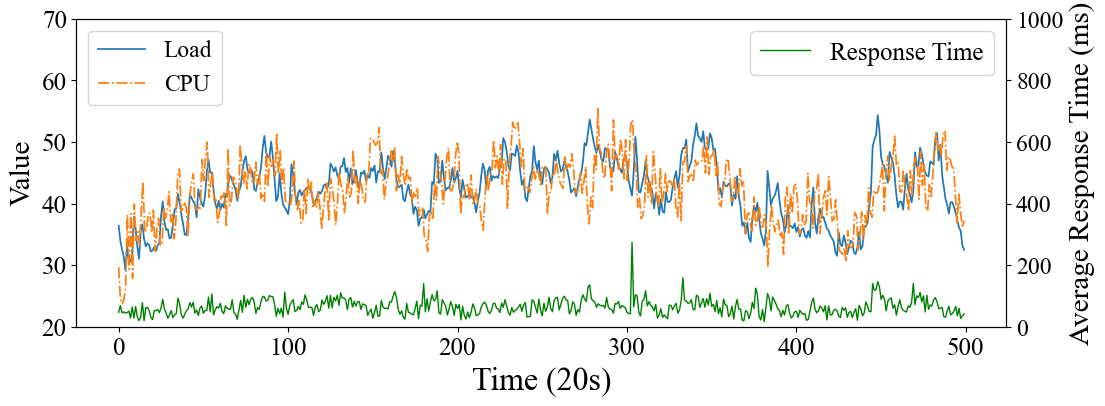}}
		\subfigure[GBMScaler]{\label{fig:subfig:b}
			\includegraphics[width=0.49\textwidth]{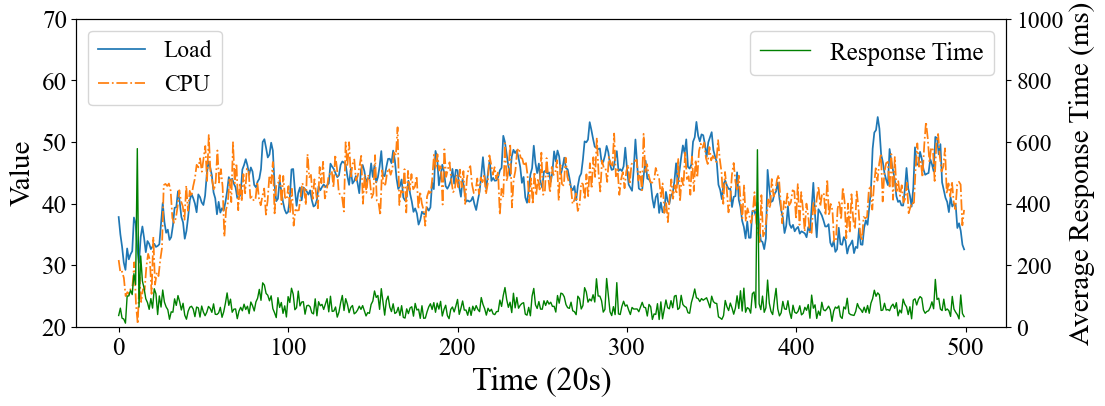}}
		\subfigure[Showar]{\label{fig:subfig:c}
			\includegraphics[width=0.49\textwidth]{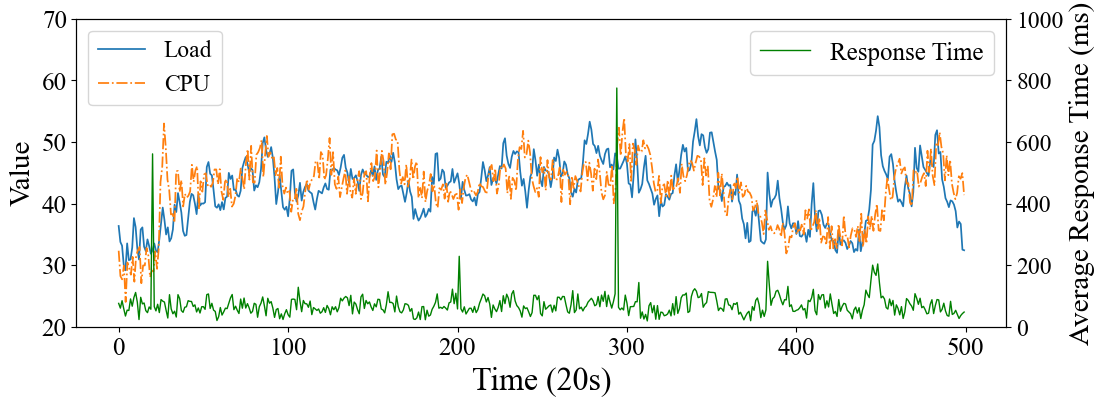}}
		\subfigure[Hyscale]{\label{fig:subfig:d}
			\includegraphics[width=0.49\textwidth]{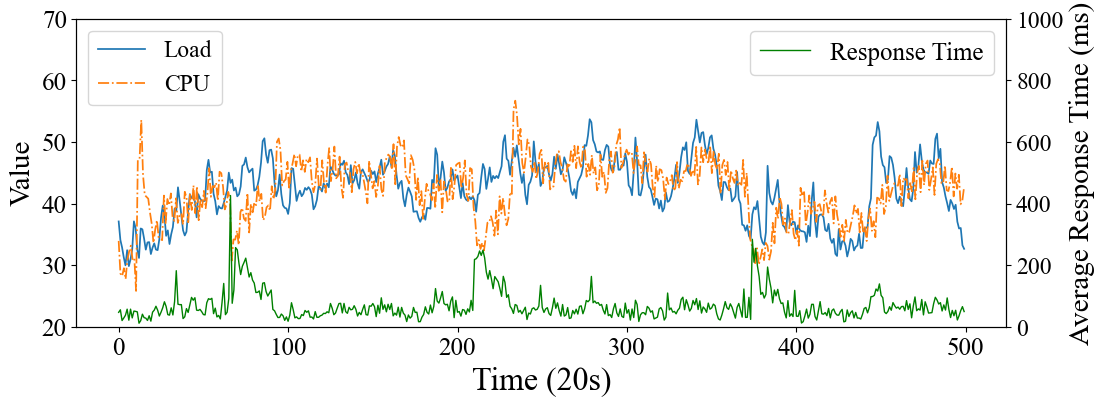}}
		\caption{	Similarity analysis with Sock-Shop application.}
		\label{fig01}
	\end{figure}
    \begin{table}
    \centering
    \caption{	Results with 95\% CI of the experiment data based on Sock-Shop.}
    \label{tab:experimental data1}
    \resizebox{\linewidth}{!}{%
    
    \begin{tblr}{
      cells = {c},
      hline{1-2,6} = {-}{},
    }
    {Elastic\\Scaling Strategy} & {Average\\Response Time (ms)} & {99th Percentile\\Response time (ms)} & {Maximum\\Response Time (ms)} & {SLO (200 ms)\\Violation (\%)} & {SLO (250 ms)\\Violation (\%)} & {Correlation\\Factor} & {Under-Provisioning\\Accuracy (\%)} & {Over-provisioning\\Accuracy (\%)}\\
    StatuScale & \textbf{63.5 (59.7, 67.4)} & \textbf{311.2 (299.6, 322.7)} & \textbf{426.1 (410.8, 441.4)} & \textbf{6.5 (5.7, 7.3)} & \textbf{3.2 (2.3, 4.1)} & \textbf{0.59 (0.56, 0.62)} & \textbf{0.8 (0.7, 1.0)} & \textbf{9.1 (5.4, 12.8)}\\
    GBMScaler & 68.5 (63.9, 73.2) & 361.8 (329.8, 393.8) & 473.1 (437.4, 508.9) & 8.1 (6.9, 9.2) & 5.9 (4.4, 7.4) & 0.53 (0.47, 0.59) & 2.0 (1.1, 2.9) & 10.1 (8.1, 12.1)\\
    Showar & 72.2 (69.9, 74.5) & 348.4 (319.3, 377.5) & 474.2 (444.6, 503.9) & 8.1 (7.3, 8.9) & 4.9 (3.9, 5.9) & 0.54 (0.50, 0.58) & 1.5 (0.5, 2.6) & 9.6 (6.5, 12.7)\\
    Hyscale & 70.0 (66.3, 73.7) & 342.8 (319.6, 366.1) & 471.5 (439.9, 503.2) & 7.4 (6.5, 8.3) & 4.3 (3.5, 5.2) & 0.50 (0.44, 0.56) & 1.1 (0.3, 1.8) & \textbf{9.1 (7.8, 10.3)}
    \end{tblr}
    }
    \end{table}
     We first conduct the performance evaluations using the microservices application Sock-Shop. The average and 99th percentile response times of each approach are compared in Fig.~\ref{fig14}, this figure represents the range of the 95th percentile CI, and the subsequent figure is similar. To better highlight the differences in response time, we use Cumulative Distribution Functions (CDF) to measure. Fig.~\ref{fig15} shows that compared with GBMScaler, Showar, and Hyscale, StatuScale can effectively reduce response time. The experimental results show that for the performance at the 99th percentile of response time, StatuScale outperforms GBMScaler by 13.99\%, Showar by 10.69\%, and Hyscale by 9.24\%. For the the performance at average response time, StatuScale outperforms GBMScaler by 7.30\%, Showar by 11.97\%, and Hyscale by 9.24\%. These results demonstrate that under the same resources budget, StatuScale's performance is significantly better than other baselines. We also observed that the performance of GBMScaler is relatively poor because it only makes decisions based on predicted results, and inaccurate predictions can have impacts on the overall performance of microservices. In terms of SLO violations, we configure the SLOs as 200 ms and 250 ms \cite{b5} respectively as shown in Fig.~\ref{fig16}, and StatuScale can achieve the lowest SLO violations compared with other baselines.

	 	 	\begin{figure}
		\centering
		\subfigure[Response time]{\label{fig14A}
			\includegraphics[width=0.235\textwidth]{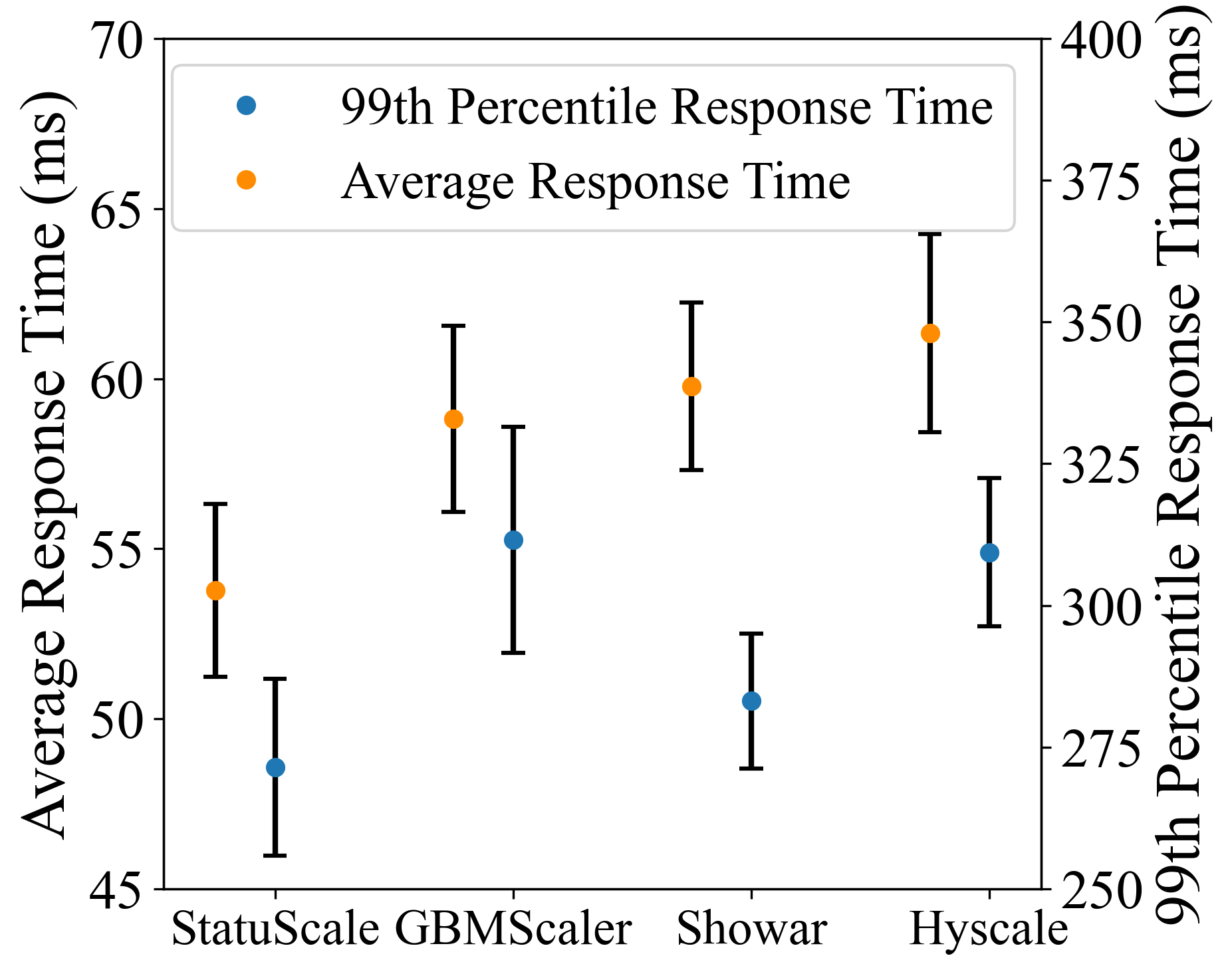}}
		\subfigure[CDF of response time]{\label{fig15A}
			\includegraphics[width=0.235\textwidth]{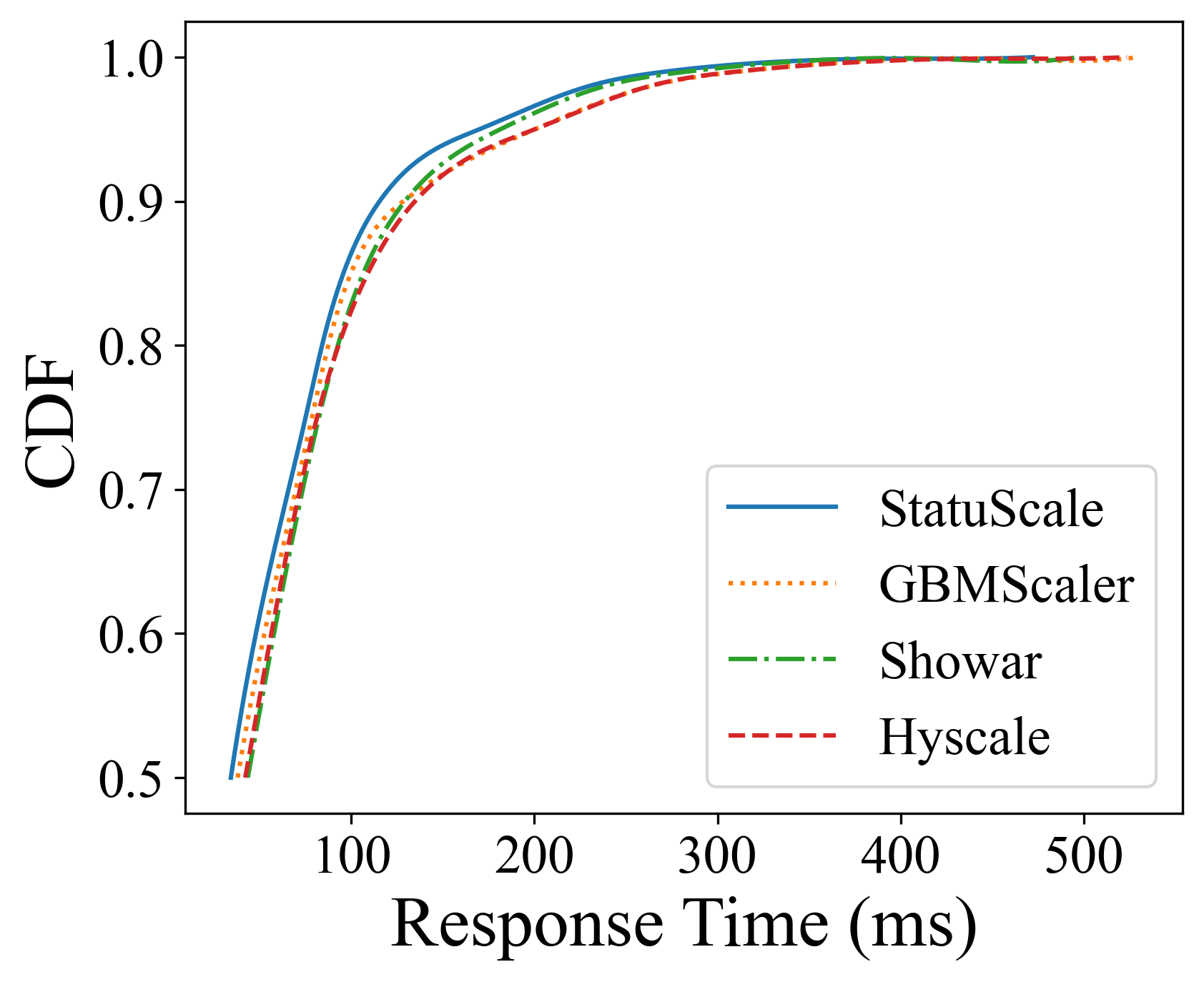}}
		\subfigure[SLO violation]{\label{fig16A}
			\includegraphics[width=0.235\textwidth]{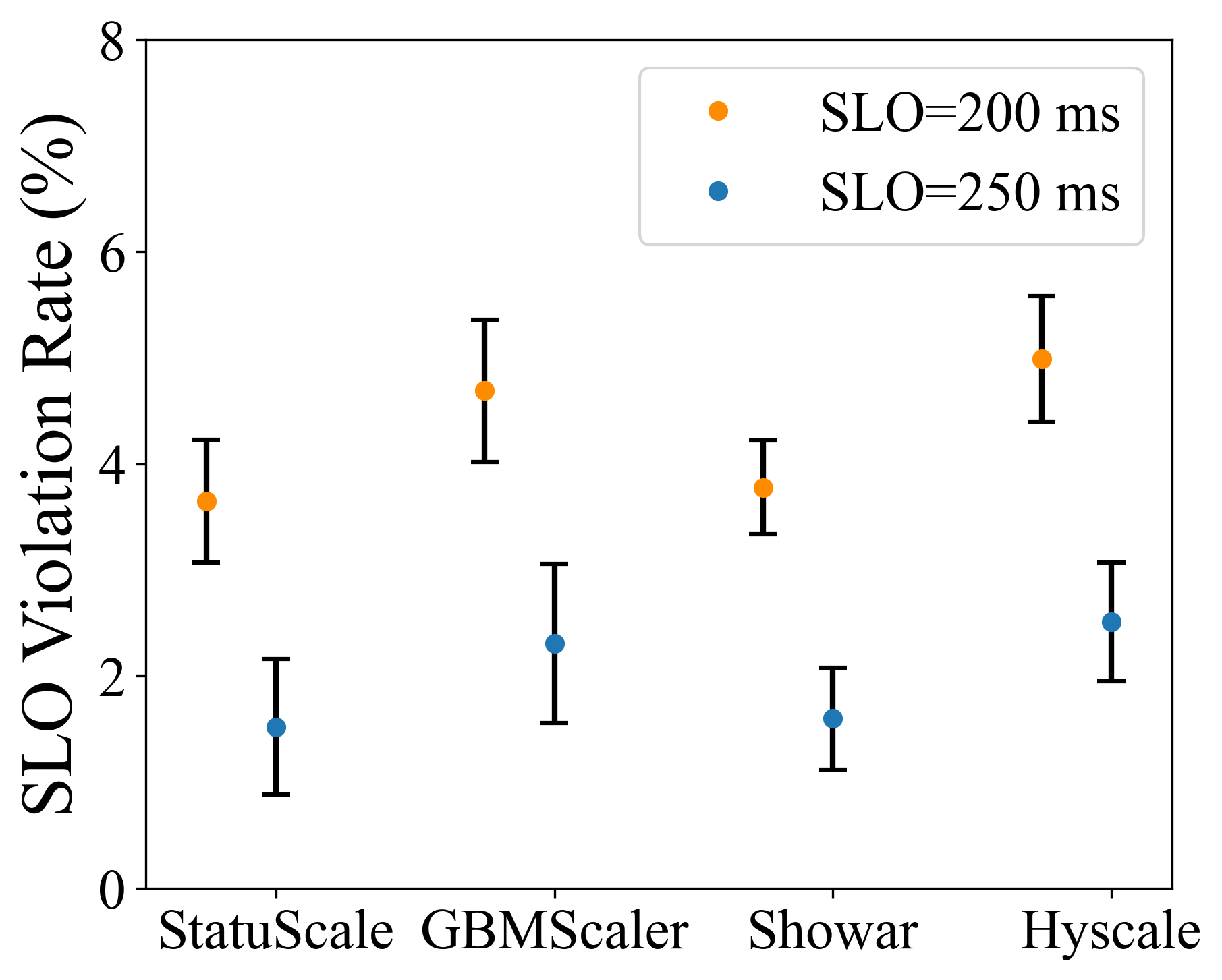}}
		\subfigure[Correlation factor]{\label{fig21A}
			\includegraphics[width=0.235\textwidth]{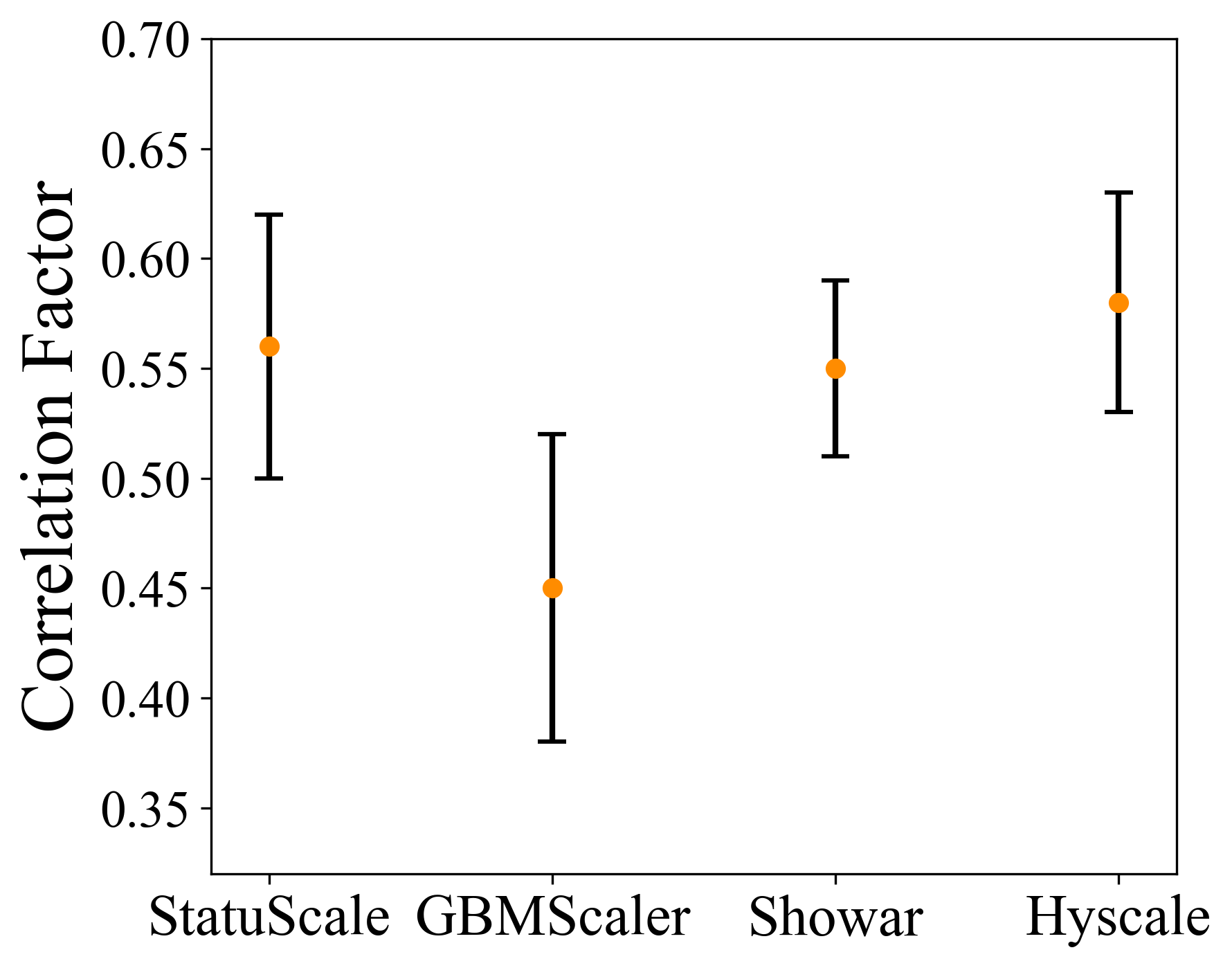}}
		\caption{	Performance comparison with Hotel-Reservation application.}
		\label{fig000}
	\end{figure}
		\begin{figure}
		\centering
		\subfigure[StatuScale]{\label{fig:subfig:a1}
			\includegraphics[width=0.49\textwidth]{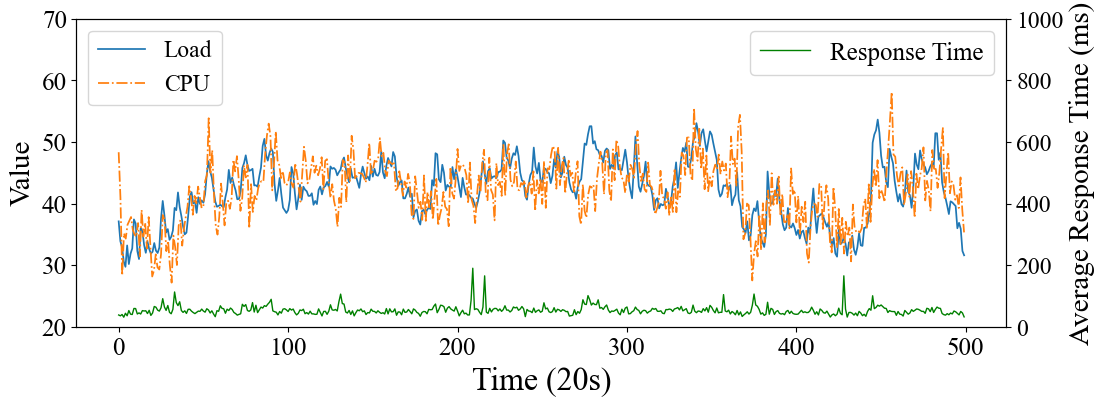}}
		\subfigure[GBMScaler]{\label{fig:subfig:b1}
			\includegraphics[width=0.49\textwidth]{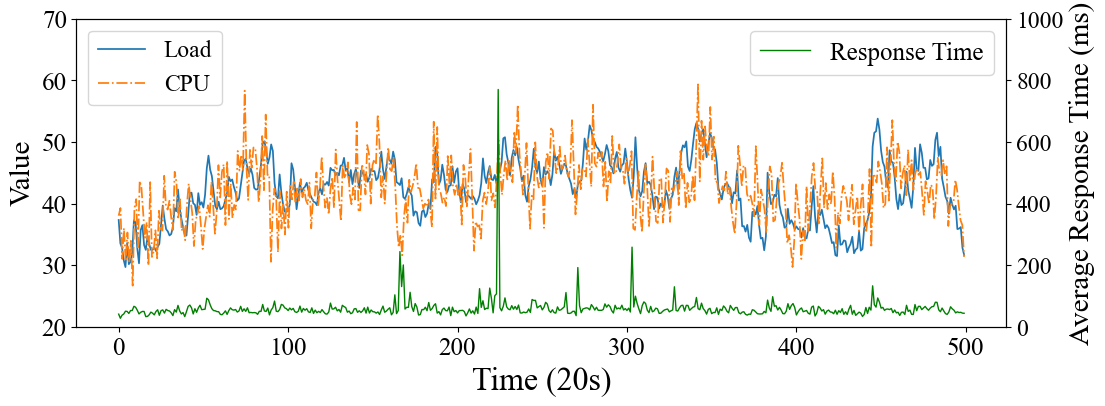}}
		\subfigure[Showar]{\label{fig:subfig:d1}
			\includegraphics[width=0.49\textwidth]{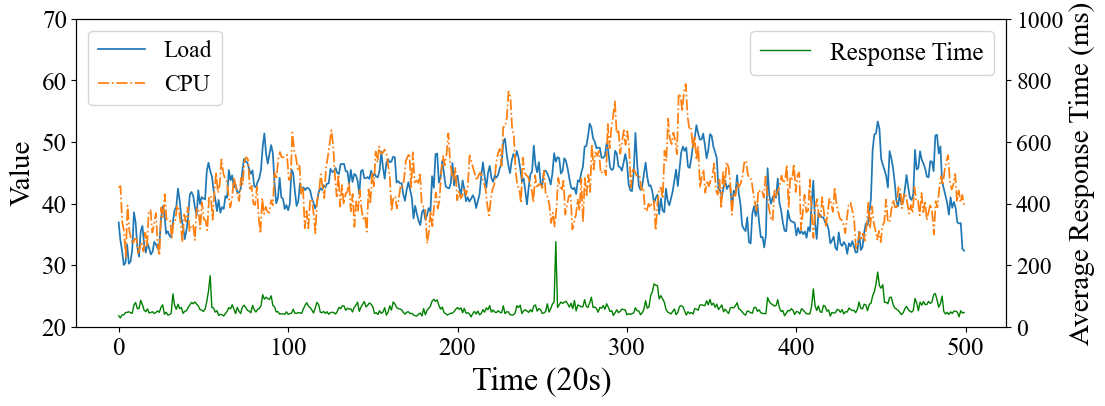}}
		\subfigure[Hyscale]{\label{fig:subfig:c1}
			\includegraphics[width=0.49\textwidth]{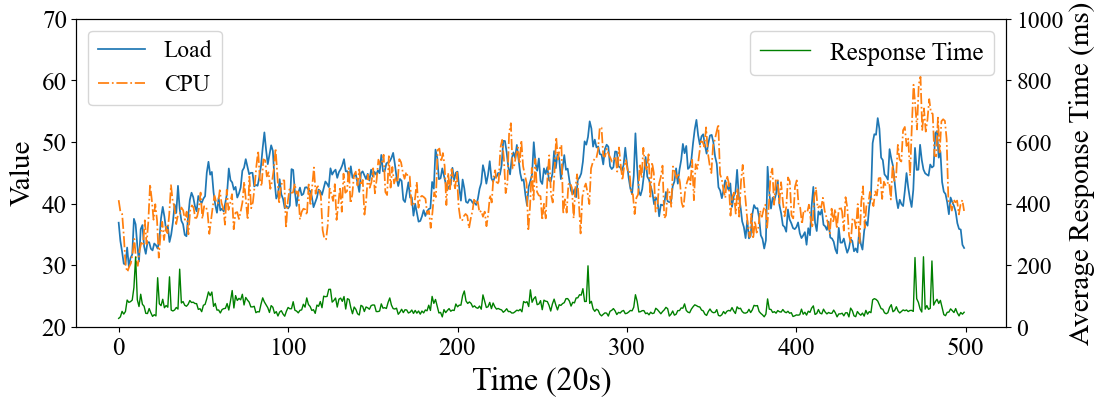}}
		\caption{	Similarity analysis with Hotel-Reservation application.}
		\label{fig:hotel_reservation}
	\end{figure}
    \begin{table}
    \centering
    \caption{	Results with 95\% CI of the experiment data based on Hotel-Reservation.}
    \label{tab:experimental data2}
    \resizebox{\linewidth}{!}{%
    
    \begin{tblr}{
      cells = {c},
      hline{1-2,6} = {-}{},
    }
    {Elastic\\Scaling Strategy} & {Average\\Response Time (ms)} & {99th Percentile\\Response time (ms)} & {Maximum\\Response Time (ms)} & {SLO (200 ms)\\Violation (\%)} & {SLO (250 ms)\\Violation (\%)} & {Correlation\\Factor} & {Under-Provisioning\\Accuracy (\%)} & {Over-provisioning\\Accuracy (\%)}\\
    StatuScale & \textbf{53.8 (51.2, 56.3)} & \textbf{271.5 (255.9, 287.0)} & \textbf{472.1 (453.5, 490.7)} & \textbf{3.7 (3.1, 4.2)} & \textbf{1.5 (0.9, 2.2)} & 0.56 (0.50, 0.62) & 0.6 (0.4, 0.7) & \textbf{17.7 (13.5, 21.8)}\\
    GBMScaler & 58.8 (56.1, 61.6) & 311.6 (291.6, 331.5) & 528.2 (504.5, 551.9) & 4.7 (4.0, 5.4) & 2.3 (1.6, 3.1) & 0.45 (0.38, 0.52) & 0.5 (0.3, 0.7) & 18.0 (13.0, 23.1)\\
    Showar & 59.8 (57.3, 62.2) & 283.2 (271.3, 295.1) & 494.7 (481.3, 508.0) & 3.8 (3.3, 4.2) & 1.6 (1.1, 2.1) & 0.55 (0.51, 0.59) & 1.1 (0.8, 1.4) & 20.2 (11.2, 29.3)\\
    Hyscale & 61.3 (58.4, 64.3) & 309.4 (296.3, 322.5) & 526.2 (510.7, 541.8) & 5.0 (4.4, 5.6) & 2.5 (2.0, 3.1) & \textbf{0.58 (0.53, 0.63)} & \textbf{0.4 (0.2, 0.5)} & 18.6 (13.1, 24.1)
    \end{tblr}
    }
    \end{table}

	To further evaluate the performance in resource elasticity scaling, we collected data on the variation of workload (requests per second) and CPU utilization over time as shown in Fig.\ref{fig01}. We then use the correlation factor to evaluate their similarity. The similarity results are shown in Fig.\ref{fig21}, StatuScale scores 0.59, GBMScaler scores 0.53, Showar scores 0.54, and Hyscale scores 0.50. StatuScale outperforms the other three baselines, demonstrating its effectiveness in resource elasticity scaling to fit with workload fluctuations. Finally, we present the relevant experimental results based on Sock-Shop in Table~\ref{tab:experimental data1}, including various evaluation metrics mentioned above. The experimental results are presented in the form of the average and 95\% CI. The optimal values are highlighted using bold formatting.

	Furthermore, we also utilized Hotel-Reservation application for evaluations. Similarly, we  compared the response time and SLO violations as shown in Fig.~\ref{fig14A}. Under the same resource budget, StatuScale outperforms GBMScaler by 8.59\%, Showar by 10.05\%, and Hyscale by 12.34\% in average response time. In terms of performance at the 99th percentile response time, StatuScale outperforms GBMScaler by 12.87\%, Showar by 4.13\%, and Hyscale by 12.26\%.

	As for SLO violations comparison, we configured SLOs threshold as 200 ms and 250 ms, as shown in  Fig.~\ref{fig16A}, StatuScale can also outperform all the baselines by reducing SLO violations. Fig. \ref{fig:hotel_reservation} shows that StatuScale can fit the resource usage and loads in a good way, and the correlation factor comparison is shown in Fig.\ref{fig21A}. Meanwhile, we present the relevant experimental results based on Hotel-Reservation in Table~\ref{tab:experimental data2}, including metrics such as under-provisioning accuracy and over-provisioning accuracy.

	\subsection{Module Evaluations}

	To further assess the effectiveness of the key component, the load status detector, in StatuScale, and evaluate the impact of each scaling mode on performance improvement, we conducted two experiments:
	
	\subsubsection{Evaluations of Status Detector Module.}
	
    In vertical scaling, the status detector module selects the appropriate elastic scaling method based on the assessed load status. This involves elastic scaling based on the LightGBM model and elastic scaling based on the A-PID controller.

	To assess the status detector module, we conducted an ablation experiment in which the module was removed. Additionally, to ensure a more precise evaluation and eliminate other interferences, horizontal scaling was also removed (the following experiments do not involve horizontal scaling). In the experiments, StatuScale with complete vertical scaling functionality (marked as StatuScale$^\triangle$), StatuScale with the load status detector and A-PID controller removed (marked as StatuScale$^{\diamond}$), and StatuScale with the load status detector and load prediction functionality removed (denoted as StatuScale$^{\ast}$) were set as our comparative algorithms.

    \begin{table}[H]
    \centering
    \caption{Evaluation results with 95\% CI of status detector module.}
    \label{vertical result}
    \resizebox{\linewidth}{!}{%
    
    \begin{tblr}{
      cells = {c},
      hline{1,5} = {-}{0.08em},
      hline{2} = {-}{},
    }
    Mertrics & StatuScale$^\triangle$ & StatuScale$^\diamond$ & StatuScale$^\ast$\\
    Average Response Time (ms) & \textbf{63.8 (60.2, 67.3)~ } & 71.5 (69.8, 73.2)~ & 67.1 (62.7, 71.5)~\\
    99th Percentile Response Time (ms) & \textbf{298.2 (292.3, 304.0)~ } & 322.4 (288.7, 356.1)~ & 315.7 (299.8, 331.5)~\\
    Max Response Time (ms) & \textbf{409.1 (402.6, 415.6)~ } & 439.4 (396.2, 482.5)~ & 429.7 (416.3, 443.0)~
    \end{tblr}
    }
    \end{table}
	Each experiment was repeated three times under essentially consistent total resource budget ($\pm 1\%$) based on microservices application Sock-Shop. We collected average response time, 99th percentile response time, and maximum response time based on each method. The average values and 95\% CI are presented in Table~\ref{vertical result}, and we marked the best result among the values in bold.
	
	The experimental results indicate that, upon removing the load status detector module, there is a noticeable decline in performance. The average response time increases between 5.24\% and 12.11\%, the 99th percentile response time increases between 5.86\% and 8.13\%, and the maximum response time increases between 5.03\% and 7.40\%. However, in methods retaining the load status detector module, experimental performance reaches its optimum.

    Additionally, it can be observed that the removal of the load status detector has a significant impact on methods based on only predictive algorithms. This is because only predictive methods inherently cannot anticipate moments of potential load underestimation, but the load status detector can achieve this.
	
	StatuScale can achieve the optimization because it accurately detects the load status and addresses resource usage issues before load bursts occur, thereby maintaining resource utilization at a stable state. Hence, it outperforms other baselines in terms of response time, SLO violations and correlation factor.
	
	\subsubsection{Evaluations of Different Scaling Modes.}
	
	As shown in Table~\ref{tab:comprison_table}, some researches focus on vertical scaling \cite{b22,b8,MaintainingSLOs}, which allows for rapidly increasing or decreasing resource supply according to demand. However, it suffers from limited scalability, as the system cannot vertically scale further once it reaches hardware limits. This makes it unsuitable for handling heavy workload, and the issue of single point of failure also restricts its widespread applicability in real-world scenarios.
	
	Conversely, some researches solely concentrate on horizontal scaling \cite{AAAI,b6,mape}. While they can offer better scalability and availability, they suffer from the issue of resource wastage, especially during periods of low workload. Even with minimal demand, a complete replica is required to handle requests.
    
    Specifically, we designed an experiment to illustrate this issue. We ran StatuScale along with its two variants to compare their resource overhead and performance differences. The first variant only utilized the vertical scaling feature of StatuScale (marked as StatuScale$^\square$), and the second variant only utilized the horizontal scaling feature of StatuScale (marked as StatuScale$^\circ$). We conducted experiments using the microservice Sock-Shop and repeated each experiment three times.

    \begin{table}[H]
    \centering
    \caption{Evaluation results with 95\% CI of different scaling modes.}
    \label{modes result}
    \resizebox{\linewidth}{!}{%
    
    \begin{tblr}{
      cells = {c},
      hline{1,6} = {-}{0.08em},
      hline{2} = {-}{},
    }
    Mertrics & StatuScale & StatuScale$^\square$ & StatuScale$^\circ$\\
    Average Response Time (ms) & 64.0 (53.6, 74.3)~ & 185.1 (183.4, 186.8)~ & \textbf{39.0 (37.0, 41.0)}~\\
    99th Percentile Response Time (ms) & 309.8 (279.2, 340.3)~ & 456.4 (450.6, 462.1)~ & \textbf{225.4 (176.0, 274.8)}~\\
    Max Response Time (ms) & 424.0 (382.5, 465.6)~ & 610.3 (605.3, 615.3)~ & \textbf{367.8 (346.4, 389.1)}~\\
    CPU Utilization (\%) & 79.5 (77.7, 81.2) & \textbf{87.4 (87.2, 87.6)} & 56.5 (54.1, 59.0)
    \end{tblr}
    }
    \end{table}

    As shown in Table~\ref{modes result}, we reported the averages values and 95\% CI of the three experiments. The experimental results indicate that StatuScale maintains an average response time of 64.0 ms with a resource utilization rate of 79.5\%. StatuScale$^\square$, with a resource utilization rate of 87.5\%, maintains the average response time at 190.5 ms. Meanwhile, StatuScale$^\circ$, with a resource utilization rate of 47.7\%, maintains the average response time at 39.6 ms. Although StatuScale$^\circ$ achieves optimal performance, its resource utilization rate is relatively low due to allocating all resources each time a complete replica is scaled, even when resource demand is low. While horizontal scaling can provide a significant increase in resources, its scaling is coarse-grained. Despite achieving the highest resource utilization rate, StatuScale$^\square$ exhibits poorer performance because of limited resources on single-machine. Once the limit of resources per machine is exceeded, vertical scaling becomes ineffective, leading to performance degradation. Overall, StatuScale, combining both horizontal and vertical scaling, offers a balanced consideration between performance and resources, making it the optimal choice for elastic scaling systems.
    
	\section{Conclusions and Future Work}\label{sec:Conclusions}
	
	This paper proposes an efficient Kubernetes-based resource management framework called StatuScale for microservices. StatuScale introduces resistance and support lines in vertical scaling, enabling differentiated resource scheduling based on load status detector. Additionally, it employs a horizontal scaling controller that utilizes comprehensive evaluation and resource reduction to manage the number of replicas for each microservice.
	
	Experiments were conducted using the typical microservice applications (Sock-Shop and Hotel-Reservation), and realistic trace derived from Alibaba. In addition, a new metric, correlation factor showing the fitness between resource usage and loads, has been used for evaluations. Results have shown that StatuScale can achieve better performance than the state-of-the-art approaches. 
	
	Despite its  advantages over the baselines, StatuScale can be further improved. For instance, the framework is built on top of individual microservices and does not fully consider the call dependency graph of microservices, studying such patterns can support ensuring various quality attributes \cite{b11,Theweakest}. Furthermore, we can optimize our resource scheduling by predicting latency \cite{tam}. In future work, we plan to consider this feature and  enhance StatuScale's ability for handling microservices with complex dependency graph.

         
	\begin{acks}
		This work is supported by National Key R \& D Program of China (No.2021YFB3300200), National Natural Science Foundation of China (No. 62072451, 62102408, 92267105), 
		Guangdong Basic and Applied Basic Research Foundation (No. 2023B1515130002, 2024A1515010251), Guangdong Special Support Plan (No. 2021TQ06X990),
		Chinese Academy of Sciences President's International Fellowship Initiative (Grant. 2023VTC0006), 
		Shenzhen Science and Technology Program (Grant No. RCBS20210609104609044), 
		Shenzhen Basic Research Program (No. JCYJ20220818101610023, KJZD20230923113800001), 
		and Shenzhen Industrial Application Projects of undertaking the National key R \& D Program of China (No.CJGJZD20210408091600002).
	\end{acks}

	\bibliographystyle{ACM-Reference-Format}
	\bibliography{sample-base}

\end{document}